\documentclass[useAMS,usenatbib]{mn2e}

\usepackage[ansinew]{inputenc}
\usepackage{graphicx}
\usepackage{pdflscape}
\usepackage{setspace,supertabular}
\usepackage{longtable} 
\usepackage{color}

\def\aj{AJ}
\def\araa{ARA\&A}
\def\apj{ApJ}
\def\apjl{ApJ}
\def\apjs{ApJS}

\def\aap{A\&A}

\def\mnras{MNRAS}
\def\pasp{PASP}

\title[Spectral Synthesis of Star-forming Galaxies in the NIR]{Spectral Synthesis of Star-forming Galaxies in the Near-Infrared}
\author[Martins et al.]{Lucimara P. Martins$^{1}$\thanks{E-mail:
lucimara.martins@cruzeirodosul.edu.br}, Alberto Rodr\'{\i}guez-Ardila$^2$, Suzi Diniz$^{1,3}$,\newauthor  Rog\'erio Riffel$^{3}$ and Ronaldo de Souza$^{4}$\\
$^{1}$NAT - Universidade Cruzeiro do Sul, Rua Galvao Bueno, 868, S\~ao Paulo, SP, Brazil\\
$^{2}$Laborat\'orio Nacional de Astrof\'isica/MCT, Rua dos Estados Unidos 154, CEP~37501-064. Itajub\'a, MG, Brazil\\
$^{3}$Universidade Federal do Rio Grande do Sul - IF, Departamento de Astronomia, CP 15051, 91501-970, Porto Alegre, RS, Brasil\\
$^{4}$Instituto Astron\^omico e Geof\'isico - USP, Rua do Mat\~ao, 1226, S\~ao Paulo, SP}

\begin{document}

\date{Accepted ? December ? Received ? December ?; in original form ? October ?}

\pagerange{\pageref{firstpage}--\pageref{lastpage}} \pubyear{2009}

\maketitle

\label{firstpage}

\begin{abstract}

The near-infrared spectral region is becoming a very useful wavelength range to
detect and quantify the stellar population of galaxies. Models are developing to predict
the contribution of TP-AGB stars, that should dominate the NIR spectra of populations 0.3 to 2 Gyr old. 
When present in a given stellar population, these stars
leave unique signatures that can be used to detect them unambiguously.
However, these models have to be tested in a homogeneous database of 
star-forming galaxies, to check if the results are consistent with what 
is found from different wavelength ranges.
In this work we performed stellar population synthesis on the nuclear and extended regions
of 23 star-forming galaxies to understand how the star-formation tracers
in the near-infrared can be used in practice.
The stellar population synthesis shows that for the galaxies with strong emission in the NIR, 
there is an important fraction of young/intermediate population contributing to the 
spectra, which is probably the ionisation source in these galaxies. 
Galaxies that had no emission lines measured in the NIR were found to have 
older average ages and less contribution of young populations. 
Although the stellar population synthesis method proved to be very effective to find the
young ionising population in these galaxies, no clear correlation between these results and 
the NIR spectral indexes were found. Thus, we believe that, in practice, the use of these indexes is
still very limited due to observational limitations.

\end{abstract}

\begin{keywords}

Stars: AGB and post-AGB, Galaxies: starburst, Galaxies: stellar content, Infrared: galaxies
\end{keywords}

\section{Introduction}

The integrated spectrum of galaxies is sensitive to the mass, age, metallicity, 
dust and star formation history of their dominant stellar populations. 
Disentangling these stellar populations is important to the understanding
of their formation and evolution and the enhancement of star formation in the universe.
Star formation tracers in the optical region are nowadays considerably
well known and studied, and have been a fundamental tool to identify 
star-formation in galaxies \citep{kennicutt88, kennicutt92, worthey+97, balogh+97, gu+06}. 
However, the use of this knowledge is not always possible in the case
of very dusty galaxies or due to the presence of a luminous AGN. 
Because of these setbacks, tracers in other wavelength regions have
been searched. 

In this sense, the near-infrared region (NIR hereafter) offers an alternative to tackle this problem. 
It conveys specific information that adds
important constrains in stellar population studies. Except for extreme cases such as
ultraluminous IRAS galaxies (Goldader et al. 1995, Lan\c con et al. 1996), the dominant
continuum source is still stellar.  
The $K$-band light of stellar populations with ages between 0.3 and 2 Gyr is dominated
by one single component, namely, the thermally pulsating stars on the asymptotic giants branch
(TP-AGB) \citep{maraston05, marigo+08}. 
For populations with age larger than 3 Gyr the NIR light is 
dominated by stars on the red giant branch (RGB) \citep{origlia+93}. Their contribution stays approximately
constant over large time scales \citep{maraston05}. By isolating the signature of these 
stellar evolutionary phases, one expects to gain a better understanding of the 
properties of the integrated stellar populations. This knowledge is
of paramount importance, for example, in the study of high redshift galaxies, 
when the major star-formation has occurred.

Population synthesis models are beginning to account for these
stars in a fully consistent way. As a result, they predict prominent 
molecular bandheads in the NIR. The spectral features of highest
relevance to extragalactic studies are the ones located redward of
1~$\mu$m, which should be detectable in the integrated
spectra of populations few times 10$^8$ years old. The detection
of these bandheads would be a safe indication of the presence
of the TP-AGB stars. The most massive of these
TP-AGB stars can be very luminous in the NIR, exceeding the luminosity of the tip of the red giant branch
by several magnitudes (Melbourne et al. 2012). 
Models that neglect TP-AGB stars have been shown to over-estimate the masses
of distant galaxies by factors of two or more in comparison to models that include them 
\citep{ilbert+10}.

Models from \citet{maraston05} show that
the combination of metallic indexes using these bands can quantify
age, metallicity and even separate populations with
single bursts from the ones with continuous star formation.
However, for nearby objects, some
of these bands are located in, or very close to, 
strong telluric absorption bands, rendering their
predictions useless or strongly dependent on the
S/N or observing conditions of the spectra.

As soon as good quality single stellar population (SSP) models in the
NIR became available, the stellar population synthesis emerged as
a powerful tool to study galaxies of many different types.

For example, by fitting combinations of stellar population models of various ages and metallicities \citet{riffel+08} 
studied the inner few hundred parsecs stellar populations of 
starburst galaxies 
in the NIR.They observed spectra were best explained by stellar populations 
containing a sizable amount (20-56\% by mass) of stars $\sim$1Gyr-old in the
thermally pulsing asymptotic giant branch. 

\citet{riffel+09} applied spectral synthesis to study the differences
between the stellar populations of Seyfert 1 (Sy1) and 2 (Sy2) galaxies in
the NIR. They found that the central few hundred parsecs of the studied galaxies contain a 
substantial fraction of intermediate-age SPs with a mean metallicity near solar.
They also found that the contribution of the featureless 
continuum and young components tends to be higher in Sy~1 than in Sy~2.

\citet{martins+10} also used stellar population synthesis
to investigate the NIR extended spectra of NGC~1068.
They found an important contribution of a young stellar population at
$\sim$ 100 pc south of the nucleus, which might be associated with regions where
the jet encounters dense clouds, possibly inducing star formation.

However, if on one hand we have now sophisticated models that predict the 
spectra of integrated populations in the NIR, 
on the other hand, few attempts have been made to fit them
to observations in a consistent way, in order to calibrate or test these predictions. 
Much of the work in the NIR
has focused on unusual objects with either active galactic nuclei
\citep{larkin+98,alonso-herrero+00,ivanov+2000,reunanen+02,reunanen+03,riffel+09, martins+10,riffel+10,riffel+11,storchi-bergmann+12} 
or very strong star formation 
\citep{goldader+97, burston+01,dannerbauer+05,
engelbracht+98, vanzi+97, coziol+01, reunanen+07,riffel+08}. 
\citet{kotilainen+12} recently published NIR long-slit spectroscopy 
for a sample of nearby inactive spiral galaxies to study the composition of 
their NIR stellar populations. With these galaxies they created NIR HK-band 
templates spectra for low redshift spiral galaxies along the Hubble sequence.
They found a dependency between the strength of the absorption lines and the 
luminosity and/or temperature of the stars, implying that NIR spectral indices can 
be used to trace the stellar population of galaxies. Moreover, evolved red stars 
completely dominate the NIR spectra of their sample, meaning that the 
contribution from hot young stars is insignificant
in this spectral region, although such ages play an important role in other spectral regions \citep{riffelRogerio+11}.

However, to identify and quantify tracers of star formation in the NIR,
we need galaxies known to have a significant fraction of star formation.
With this in mind, we used the NIR spectral sample of star-forming
galaxies of \citet{martins+13}, which are known to have
star formation from their optical observations. 
Our objective is to test the predictions of stellar population
models in this wavelength range and 
verify the diagnostics that can,  in practice, be used as proxies in stellar
population studies. 
In order to do this we fit the underlying continuum between 0.8 and 2.4 $\micron$ 
with the stellar population synthesis technique, using the same method described in 
\citet{riffel+09} and \citet{martins+10}.
In \S 2 we present the details of our observations and reduction
process; in \S 3 we briefly describe the stellar population synthesis method; 
in \S 4 we present our results and discussions and in  \S 5 we show our conclusions.

\section{The Data}

The sample used here was presented in \citet{martins+13} and
is a subset of the one presented in the magnitude-limited
optical spectroscopic survey of nearby bright galaxies of Ho et al. (1995, hereafter HO95). 
These galaxies are sources defined by Ho et al. (1997, hereafter HO97) as those composed 
of ``nuclei dominated by emission lines from regions of active star formation (H\,{\sc ii} or starburst
nuclei)". In addition, five galaxies classified as non-star-forming
in the optical, dominated by old-stellar population and with no detected emission lines, 
were included as a control sample. 
 
All spectra were obtained at the NASA 3m Infrared
Telescope Facility (IRTF) in two observing runs (2007 and 2008) - the same 
data from \citet{martins+13}. The
SpeX spectrograph \citep{rayner+03} was used in the short
cross-dispersed mode (SXD, 0.8-2.4 $\mu$m).  The detector
consists of a 1024x1024 ALADDIN~3 InSb array with a spatial scale
of 0.15"/pixel. A 0.8"x 15" slit oriented in the north-south direction
was used, providing a spectral resolution of 360 km\,s$^{-1}$. This value was 
determined both from the arc lamp and the sky line spectra and was found to
be constant with wavelength along the observed spectra. The seeing
varied from night to night but on average, most objects were observed
under a 1$\arcsec$ seeing conditions.

The spectral reduction, extraction and wavelength calibration
procedures were performed using SPEXTOOL, the in-house
software developed and provided by the SpeX team for
the IRTF community \citep{cushing+04}. Telluric
features removal and flux calibration were done using XTELLCOR \citep{vacca+03},
another software available by the SpeX team. The spectra were corrected
for Galactic extinction using the \citet{cardelli+89} law 
and \citet{schlafly+11} extinction map.
 
The final sample is composed of 28 galaxies, from which 23 are classified 
as starbursts or star-forming and 5 are non-starforming galaxies for comparison. 
A different number of apertures were extracted for each galaxy, depending on 
the size of the extended emission across the slit. 
The average exctraction size is 1". More details about the 
observations and reduction process can be found in \citet{martins+13}.
Additional information about these galaxies are presented in Table~\ref{sample}.
The G band measurements and the H$_\alpha$ flux are presented because
they can be used to characterize the old and young stellar populations,
respectively, in the optical.

The wavelength and flux-calibrated spectra were de-redshifted using the
$z$ values listed in column~2 of Table~\ref{sample}. Then, they were normalised
to unity at 1.233\,$\mu$m (as defined in \citet{riffel+08}). This position was set because no emission 
or absorption features are located in or around it.

\begin{table}
\centering
\caption{ Sample details}
\footnotesize
\setlength{\tabcolsep}{3pt}
\begin{tabular}{@{}ccccccccccc@{}}
\hline
Galaxy &     z  &   Apertures & Class & W{\tiny (G band)}$^*$ & log F(H$_\alpha$)$^*$\\
       &        &      (a)    & (b)   & (c)&  (d)                     \\
\hline

NGC 0221  &-0.0007  &   nuc + 3 & N   & 4.54 & -       \\
NGC 0278  & 0.0021  &   nuc + 3 & H   & 0.50 & -13.45  \\
NGC 0514  & 0.0082  &   nuc + 2 & H   & 3.86 & -14.64  \\
NGC 0674  & 0.0104  &   nuc + 2 & H   & 4.38 & -14.31  \\
NGC 0783  & 0.0173  &   nuc + 2 & H   & 1.98 & -13.54  \\
NGC 0864  & 0.0052  &   nuc + 2 & H   & 1.03 & -12.88  \\
NGC 1174  & 0.0091  &   nuc + 2 & H   & 3.04 & -13.07  \\
NGC 1232  & 0.0053  &   nuc     & N   &   -  & -       \\
NGC 1482  & 0.0064  &   nuc +2  & H   &   -  & -       \\
NGC 2339  & 0.0074  &   nuc + 2 & H   & 0.00 & -13.07  \\
NGC 2342  & 0.0176  &   nuc     & H   & 0.67 &  -12.83 \\
NGC 2903  & 0.0019  &   nuc + 7 & H   & 0.57 &  -12.63\\ 
NGC 2950  & 0.0045  &   nuc + 4 & N   & 4.98 &  -	\\
NGC 2964  & 0.0044  &   nuc + 2 & H   & 1.03 &  -12.75 \\
NGC 3184  & 0.0020  &   nuc + 2 & H   & 1.37 &  -13.12 \\
NGC 4102  & 0.0028  &   nuc + 2 & H   & 1.94 &  -12.52\\
NGC 4179  & 0.0042  &   nuc + 2 & N   & 4.89 &  -14.30\\
NGC 4303  & 0.0052  &   nuc + 6 & H   & 3.60 &  -12.84\\
NGC 4461  & 0.0064  &   nuc + 2 & N   & 4.67 &  -     \\
NGC 4845  & 0.0041  &   nuc + 6 & H   & 4.18 & -13.61 \\
NGC 5457  & 0.0008  &   nuc + 2 & H   & 1.03 & -13.33 \\
NGC 5905  & 0.0113 	&   nuc + 2 & H   & 2.15 & -13.13 \\
NGC 6181  & 0.0079 	&   nuc + 2 & H   & 4.43 & -13.57 \\
NGC 6946  & 0.0195  &   nuc     & H   & 0.32 & -13.01 \\
NGC 7080  & 0.0161	&   nuc + 2 & H   & -    & -13.49 \\
NGC 7448  & 0.0073 	&   nuc + 2 & H   & 0.42 & -14.01\\
NGC 7798  & 0.0080  &   nuc + 2 & H   & 0.04 & -13.16\\
NGC 7817  & 0.0077	&   nuc + 2     & H   & 3.57 & -13.25\\

\hline
\end{tabular}
\raggedright
\\
References:
(1) \citet{ho+95}
(2) \citet{kennicutt88}
(3) \citet{coziol+98}
\\
(*)Adopted from \citet{ho+97}.//
Redshifts obtained from NED (http://ned.ipac.caltech.edu/) 
(a) Number of apertures extracted from each galaxy. (b) H means star-forming and 
N means non-star-forming galaxy. (c) Measured G band. 
(d) Log of the measured H$\alpha$ flux (in ergs/cm$^2$/s) 
\label{sample}
\end{table}

\section{Spectral Synthesis Base Set and Method}

The spectral synthesis is done using the code STARLIGHT
\footnote{The code and its manual can be downloaded from http://astro.ufsc.br/starlight/node/1}
\citep{cid+04,cid+05a,mateus+06, asari+07,cid+09}, 
which mixes computational techniques
originally developed for semi-empirical population synthesis with 
ingredients from evolutionary synthesis models.
The code fits an observed spectrum O$_\lambda$ with a combination,
in different proportions, of a number SSPs,
to obtain the final model that fits the spectrum, M$_\lambda$.
Due to the fact that the \citet{maraston05} models include the effect of the TP-AGB
phase, we use these SSPs models
as the base set for STARLIGHT. The SSPs used in this work cover
14 ages, t = 0.001, 0.005, 0.01, 0.03, 0.05, 0.1, 0.2, 0.5, 0.7, 1, 2, 5, 9
and 13 Gyr, and 4 metalicities, Z = 0.02, 0.5, 1 and 2 z$_{\sun}$,
summing up 56 SSPs.

It is important to mention that the spectral resolution of the models in 
this spectral region ranges from 
R $\sim$100 (K-band) to R $\sim$ 250 (z-band) while that of the observed data is significantly higher
R $\sim$750 \citep{rayner+03}. The observations were then degraded to the model's resolution. 
For this reason, STARLIGHT's fit
will depend more strongly on the continuum shape rather than on the absorption features. 
This comes directly from the way the synthesis works: the best models are
chosen based on the $\chi^2$ values. This means that the synthesis does a point by
point subtraction between models and observations. When the resolution
is very low, the ``distance'' between models and observations
becomes smaller in the absorption bands, and therefore they become
``less important''.
Recently, \citet{maraston+11} published a new set of NIR models
with R $\sim$500 (K-band). However, they are only available for solar metallicity. 
In order to test possible differences introduced in the results due
to the spectral resolution, these latter models were used to construct a base 
using the same ages as for the low resolution base.

Besides that, we include in our spectral
base 8 Planck distributions (black-body-BB), with T ranging from 
700 to 1400 K, in steps of 100 K, to account for a possible contribution
of hot dust to the continuum (for more details see \citet{riffel+09}).
Although dust at such high temperatures is not common in starburst
galaxies, its existence would indicate the presence of a hidden 
active galactic nucleus, for instance.

Extinction is modelled by STARLIGHT as due to foreground dust,
and parametrised by the V-band extinction A$_V$. We
use the \citet{calzetti+00} extinction law to this purpose
because it is more appropriated for star-forming galaxies \citep{Fishera+03}.  

Velocity dispersion is also a free parameter for STARLIGHT. This means that
the code broadens the SSPs in order to better fit the absorption lines in the
observed spectra. In our case the velocity dispersion results lack of significance 
because of the low resolution of the models compared to that of the observations.

Emission lines were masked, since STARLIGHT only fits the stellar population continuum.
Lines masked in this work were [SIII]$\lambda$ 0.907 $\micron$, [SIII]$\lambda$ 0.953 $\micron$, 
HeI 1.083 $\micron$, Pa$\gamma$ $\lambda$ 1.094 $\micron$, [FeII]$\lambda$ 1.257 $\micron$,   
Pa$\beta$ $\lambda$ 1.282, [FeII]$\lambda$ 1.320 $\micron$, [FeII]$\lambda$ 1.644 $\micron$, 
Pa$\alpha$ $\lambda$ 1.875 $\micron$ and HI $\lambda$ 1.945 $\micron$.Spurious data (like bad teluric correction regions) were also masked
out.

\section{Results}

\begin{figure*} 
\includegraphics [width=180mm]{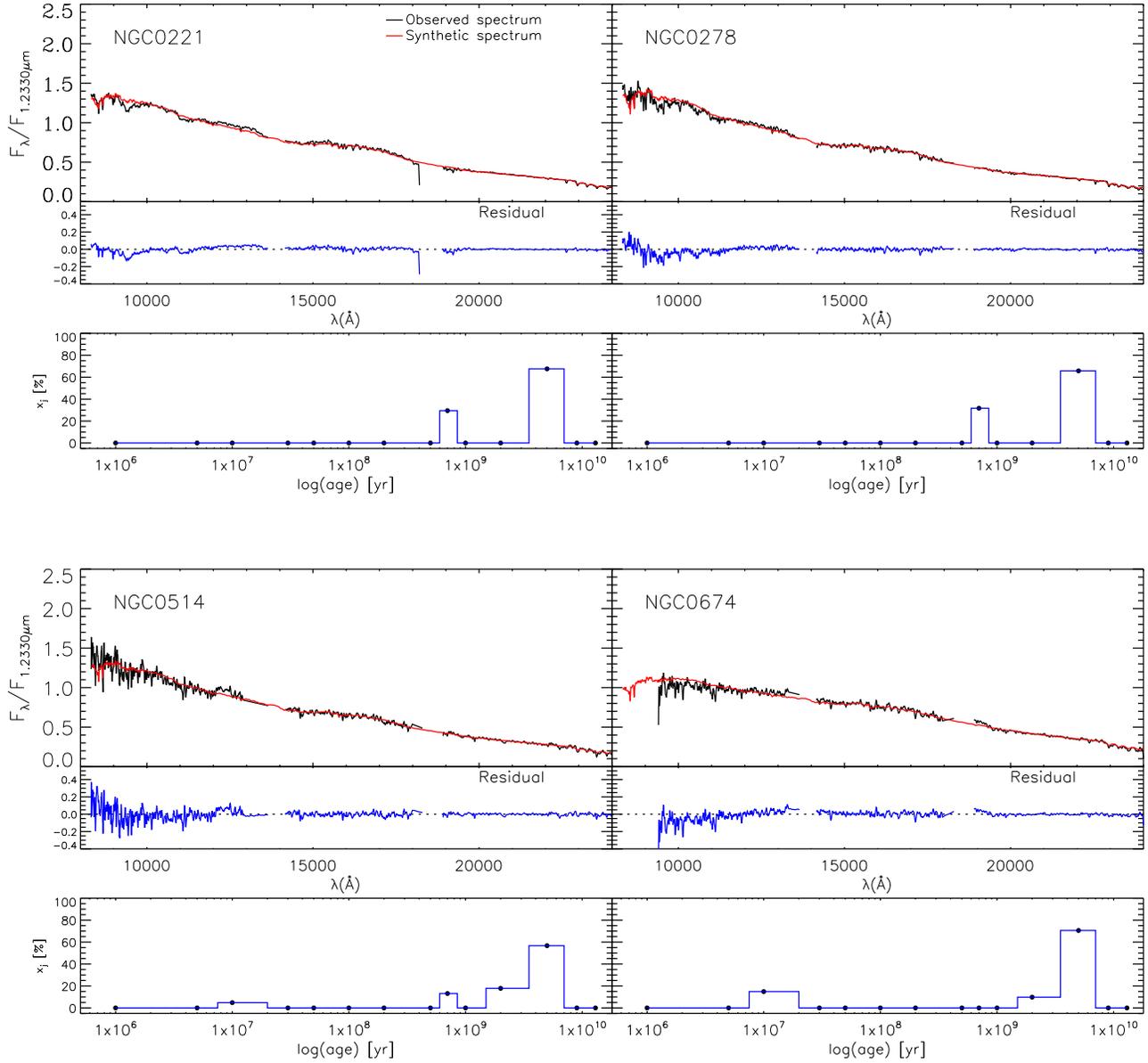}
\caption{Results of the spectral fitting procedure for the nuclear apertures of 
NGC~0221, NGC~0278, NGC~0514 and NGC~0674. Each panel shows at the top, the flux of the
observed spectrum and the synthetic spectrum found by the synthesis, normalised at unity
at 1.233 $\mu$m. The middle panel shows the residual spectrum (observed minus synthetic spectrum).
The bottom panel shows the contribution of each SSP used in the base to the continuum.}
\label{specfit1}
\end{figure*}

\begin{figure*} 
\includegraphics [width=180mm]{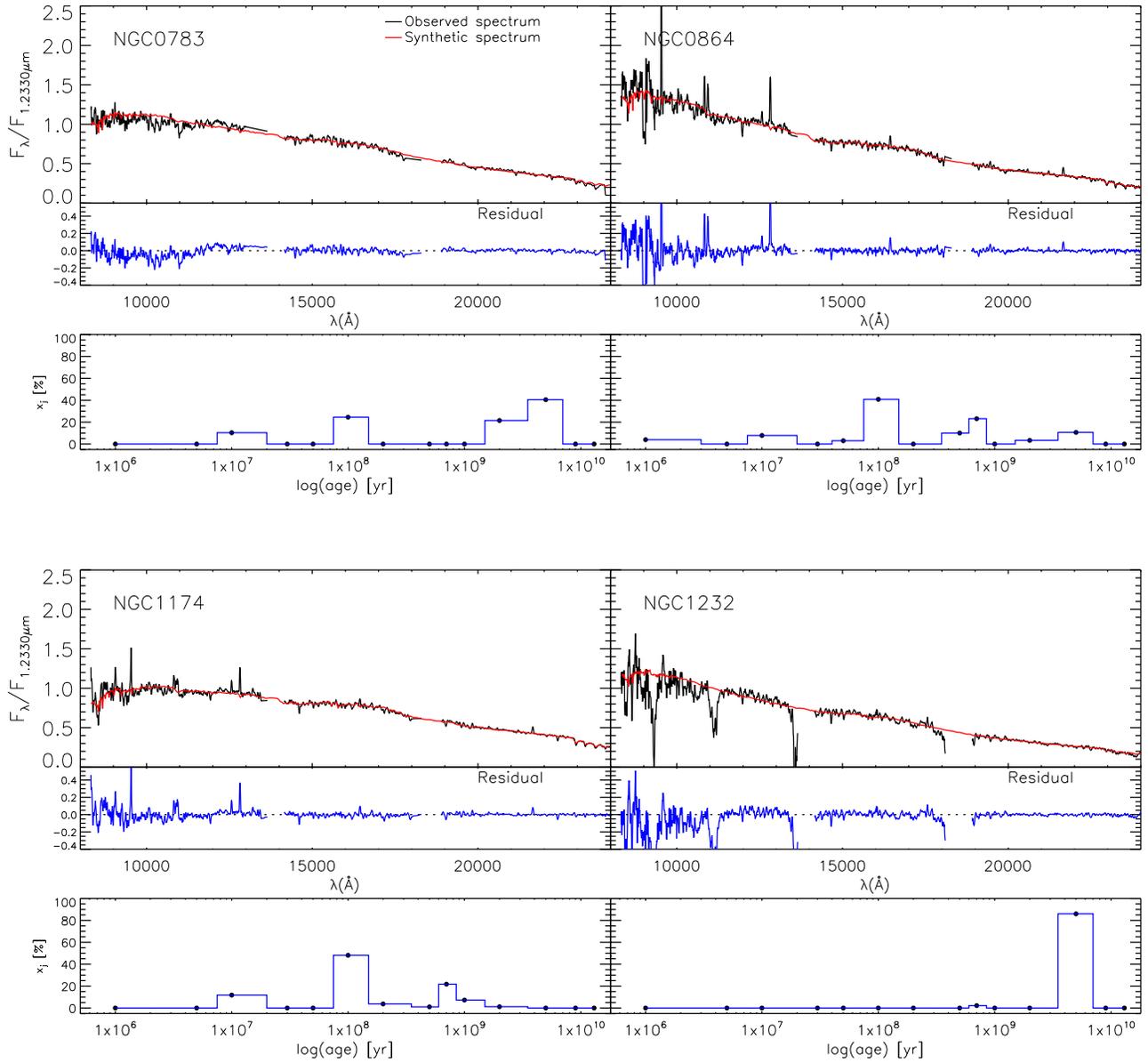}
\caption{The same as Fig.~\ref{specfit1}, but for 
NGC~0783, NGC~0864, NGC~1174 and NGC~1232.}
\label{specfit2}
\end{figure*}

\begin{figure*} 
\includegraphics [width=180mm]{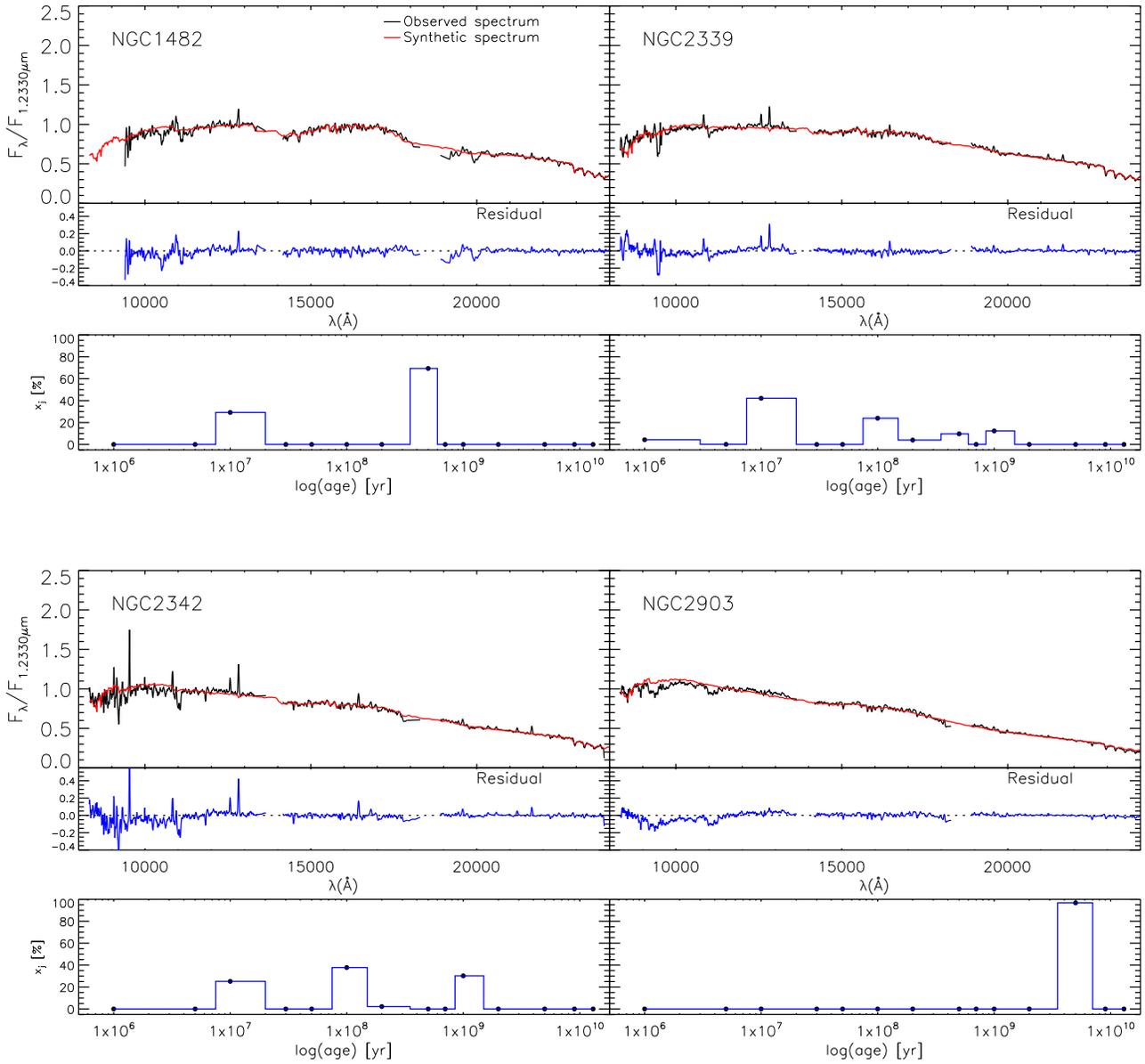}
\caption{The same as Fig.~\ref{specfit1}, but for 
NGC~1482, NGC~2339, NGC~2342 and NGC~2903.}
\label{specfit3}
\end{figure*}

\begin{figure*} 
\includegraphics [width=180mm]{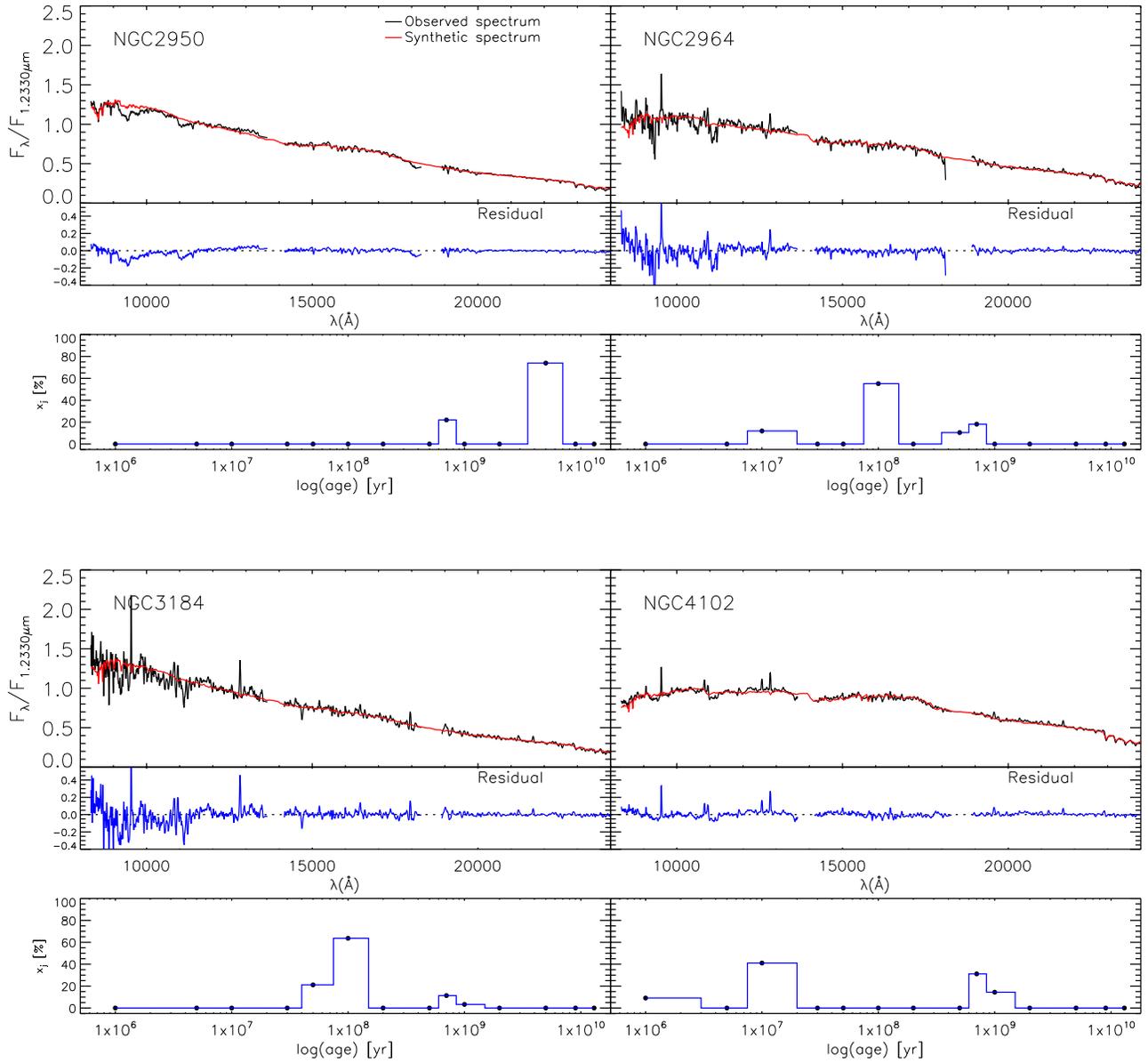}
\caption{The same as Fig.~\ref{specfit1}, but for 
NGC~2950, NGC~2964, NGC~3184 and NGC~4102.}
\label{specfit4}
\end{figure*}

\begin{figure*} 
\includegraphics [width=180mm]{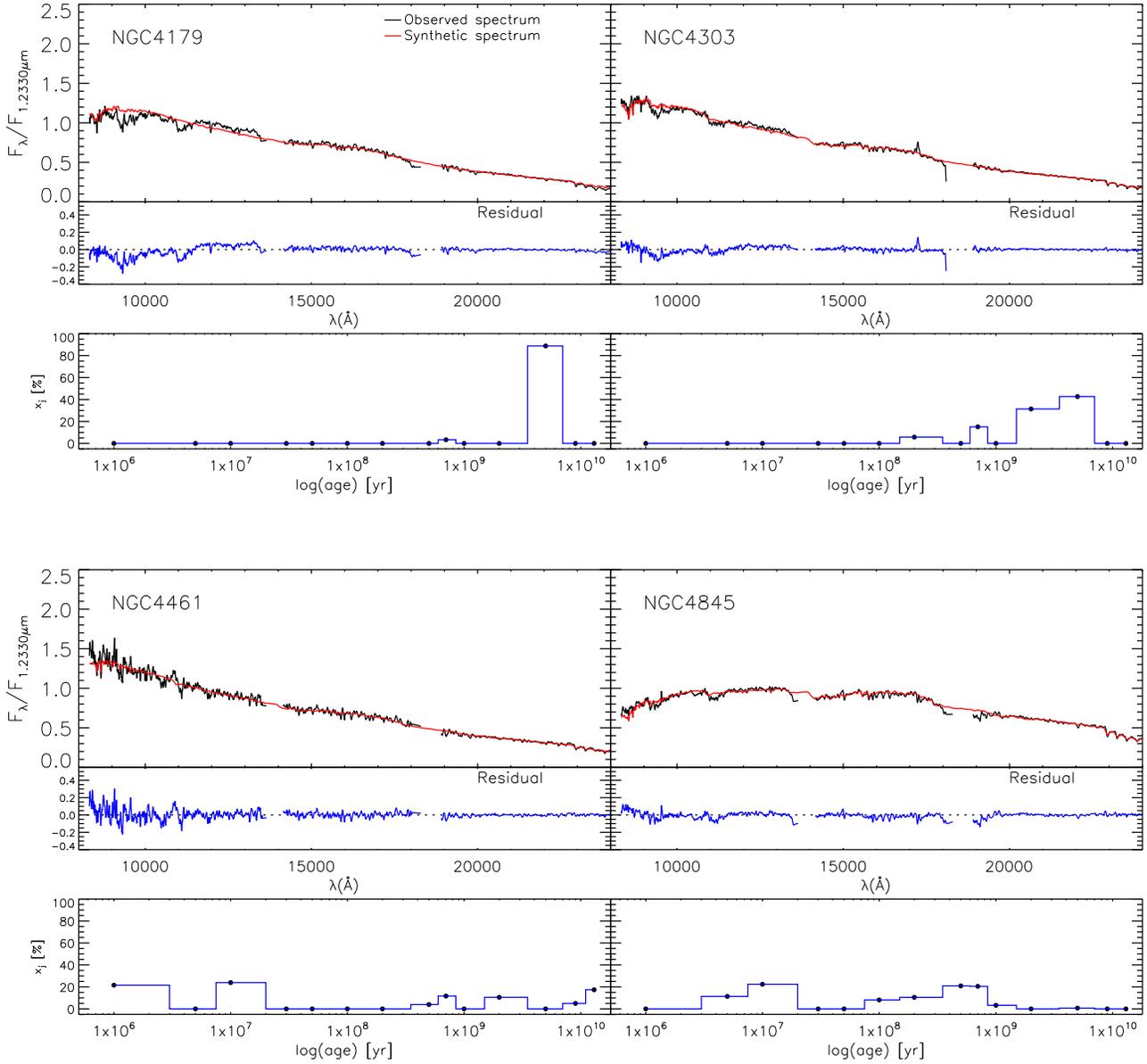}
\caption{The same as Fig.~\ref{specfit1}, but for 
NGC~4189, NGC~4303, NGC~4461 and NGC~4845.}
\label{specfit5}
\end{figure*}

\begin{figure*} 
\includegraphics [width=180mm]{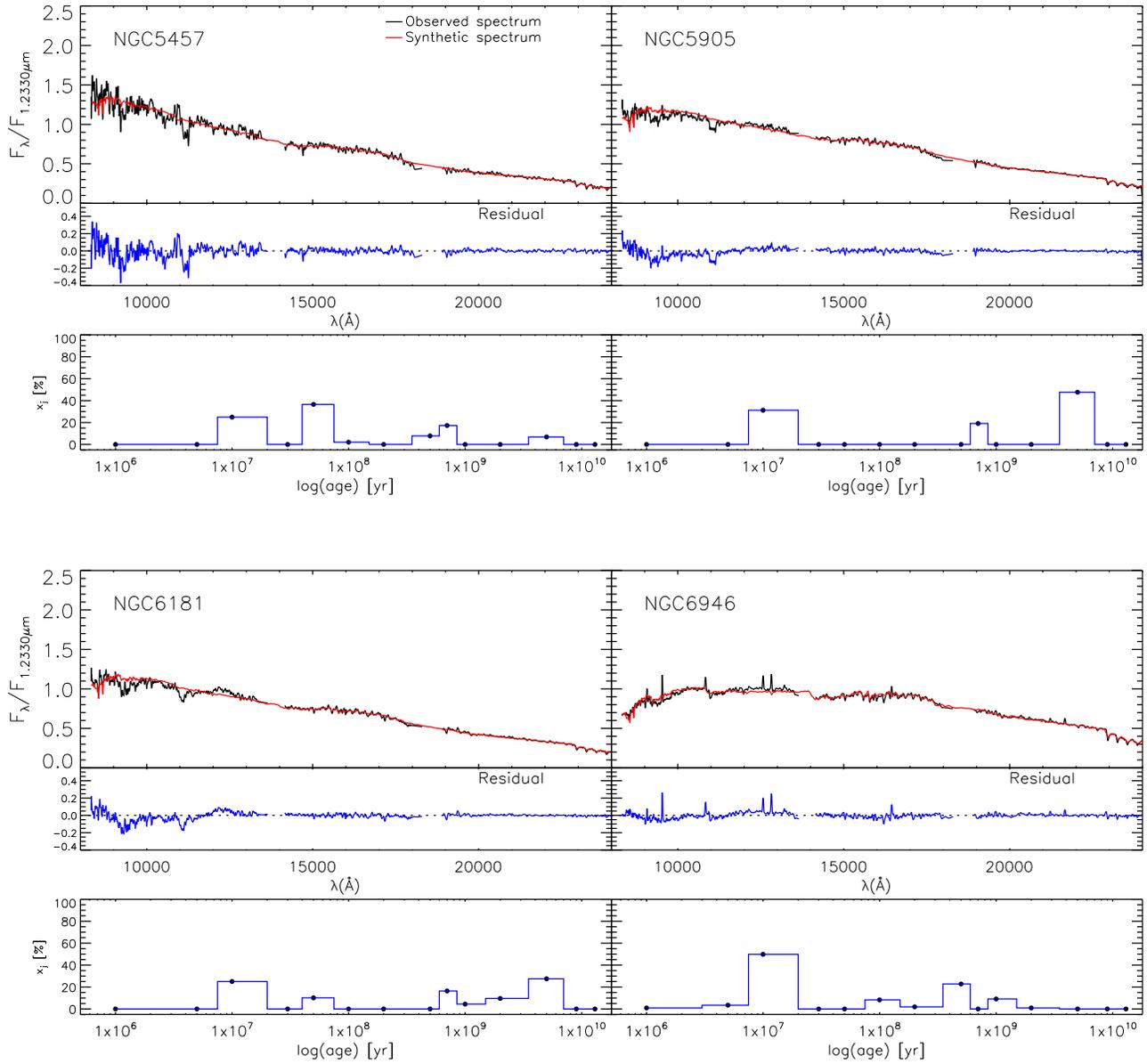}
\caption{The same as Fig.~\ref{specfit1}, but for 
NGC~5457, NGC~5905, NGC~6181 and NGC~6946.}
\label{specfit6}
\end{figure*}

\begin{figure*} 
\includegraphics [width=180mm]{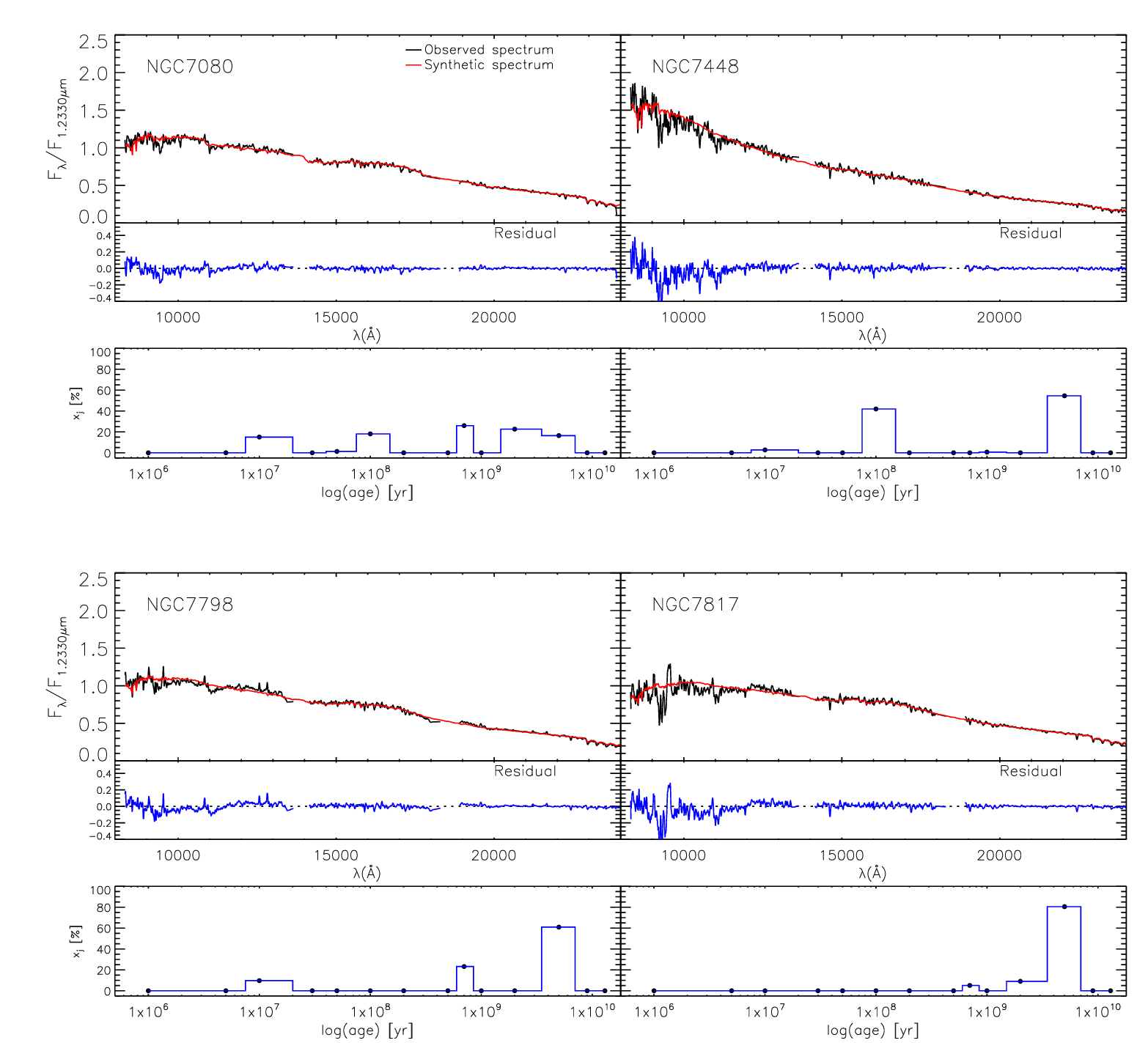}
\caption{The same as Fig.~\ref{specfit1}, but for 
NGC~7080, NGC~7748, NGC~7798 and NGC~7817.}
\label{specfit7}
\end{figure*}

The results of the spectral synthesis fitting procedure for the nuclear
apertures are presented in Figures~\ref{specfit1} to~\ref{specfit7}.   
For each galaxy, the top panel shows the observed and modeled
spectra normalised to unity at 1.233 $\mu$m. The middle panel
shows the observed spectrum after subtraction of the stellar
contribution (O$_\lambda$ - M$_\lambda$). This residual can be understood as
the gas emission component. The
bottom panel shows the contribution of each SSP
to the continuum found by the synthesis. This last panel
can be understood in terms of the star formation history of the
galaxy.

Figures~\ref{specfit1} to~\ref{specfit7} show that overall, the
the spectral synthesis reproduces well the continuum shape and the most
conspicuous absorption signatures, like the CaT and CO. 
Examples of good results are NGC~0864 and NGC~4102.
However in many galaxies
some strong absorption signatures are not being
reproduced by the models. Clear examples are the features that appear around
0.94 $\mu$m (TiO) and 1.1 $\mu$m (CN) in many galaxies.
Worst cases seem to be NGC~2950, NGC~4179 and NGC~6181. It is important to mention that we tested many different weights for different
regions in the synthesis process, but the features were still not reproduced.
Many things might be playing a role here, from telluric contamination
around these regions to lack of precision from the method due to the low resolution of 
the models, or even the strength of these signatures in the models.

The quality of the fits are measured by the reduced $\chi^2$ and the {\it adev} parameter,
which is the percentage mean deviation over all fitted pixels, 
$|$O$_\lambda$ - M$_\lambda$$|$/O$_\lambda$.
Following \citet{cid+05b}, we present our results 
using a condensed population vector to take
into account noise effects that dump small differences between similar spectral components.
This is obtained by binning the population vector x into young (x$_Y$: t $\leq$ 5
$\times$ 10$^7$ yr), intermediate-age (x$_I$: 1 $\times$ 10$^8$ $\leq$ t $\leq$ 2 $\times$
10$^9$ yr) and old (x$_O$: t $>$ 2 $\times$ 10$^9$ yr) components, using the
flux distributions. 
Results for the condensed vectors for each galaxy (nuclear and off-nuclear apertures) are presented in Table~\ref{synthesis}.
Additional results from the synthesis, namely the extinction value A$_v$,
the mean age $<$log(t$_{\textrm{av}}$)$>$ and mean metallicity $<$z$_{\textrm{av}}$$>$ of the stellar population,
weighted by the light fraction are also shown.
These quantities are defined as \citet{cid+05b}:

\begin{equation}\langle {\rm log} t_{ av} \rangle_{L} = \displaystyle \sum^{N_{\star}}_{j=1} x_j {\rm log}t_j, \end{equation}

and

\begin{equation}\langle Z_{ av} \rangle_{L} = \displaystyle \sum^{N_{\star}}_{j=1} x_j Z_j,\end{equation}


where N$^*$ is the total number of SSPs used in the base.
Table~\ref{synthesis} also presents the current mass of the galaxy which is in the form of stars (M*), the mass that was processed
in stars during the galaxy lifetime (M$_{ini}$) and the star formation rate in the last 100 Myrs (SFR$_{100}$).
We also present the  reduced $\chi^2$ and the {\it adev} values for each fit. The last column of this table shows the distance of the aperture (in arcseconds) to the nucleus.
More information about these quantities can be obtained
in the STARLIGHT manual.

\citet{cid+05b} tested the effects of the S/N in the optical on the
stellar population synthesis results, concluding that for observed spectra
with S/N $\ge$ 10, in the continuum the results are robust. For that reason we only applied
the synthesis method for apertures obeying this criteria.

Average metallicites tend to be between 1 and 2 times the solar value.
No hot dust contribution to the continuum was found from the 
synthesis for any galaxy. 
This result rules out the presence of a hidden AGN in our sample and agrees with
previous observations in other wavelengths bands, where no mention to the possible 
presence of an AGN, even of low-luminosity, is made for any of the targets studied. 
Moreover, we found a significant contribution of young stellar population (x$_Y$ $\geq$ 10\%)
in 17 out of the 23 galaxies classified as starforming galaxies.
In contrast, four of the five galaxies classified as non-star-forming have zero contribution 
of young stars, being dominated by an old stellar population whose contribution is
larger than 75\%. This agrees with the work of \citet{kotilainen+12} who
also found negligible contribution of young stars in their sample of non-active
galaxies by means of $H-$ and $K$-band spectroscopy. NGC~4461 the only non-star-forming galaxy with
anomalous result, has a young population fraction of x$_Y$ $\sim$ 48\% in the
nuclear region. This is surprising, as no emission lines were detected in that object.
Moreover, its NIR spectrum is almost featureless, with no apparent CaII or CO 2.3$\mu$m
absorption features, leading us to consider this result with caution. 
Indeed,
the remaining non-star-forming galaxies of our sample, as well as those of \citet{kotilainen+12},
display these two systems of absorption lines/band. We therefore believe that our fits 
failed because no constraint could be set either from the continuum or from absorption 
features. 

Ideally one would like to compare these results with stellar populations synthesis
results from the optical. However, as shown by \citet{martins+13}, the apertures from the optical
observations and from the NIR are very diffent, which leads to the sampling of different stellar
populations (the optical spectra were obtained through a slit with 5 times the area of the NIR slit). 
Despite this, we believe our results can be trusted as consistent with the optical emission line tracers
for the stellar population, which is further detailed in section 4.3.

\begin{figure*} 
\includegraphics [width=180mm]{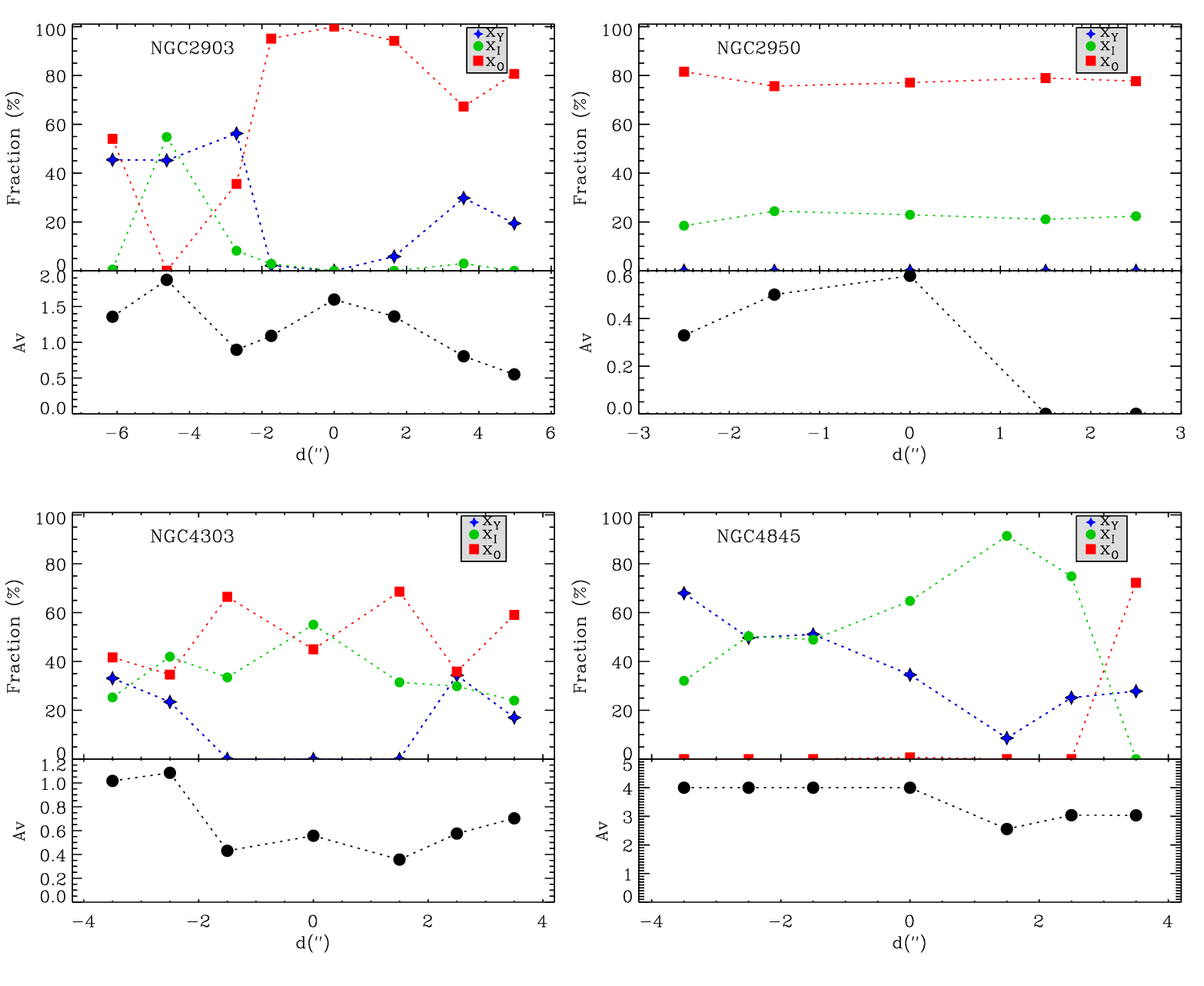}
\caption{Fractions of stellar populations obtained from the synthesis as a function of the distance to 
the centre for galaxies with five or more apertures. Negative distances represent the north direction
and positive distances represent the south direction.}
\label{gradpop}
\end{figure*}

Figure~\ref{gradpop} shows the gradient of the
stellar populations for galaxies with five or more apertures. In this plot, negative distances
represent the north direction and positive distances the south. One of these galaxies
is a non-star-forming galaxy (NGC~2950, top right plot), and no variation of the stellar population
is seen along the galaxy. For the three remaining galaxies (NGC~2903, NGC~4303 and NGC~4845), the
fraction of younger populations are clearly rising in the outer apertures, coinciding
with where the emission lines are stronger.

\onecolumn
\clearpage
\begin{center}
\small
\setlength{\tabcolsep}{5pt}
\begin{longtable}{cccccccccccccc}
\kill
\caption[]{Results of the NIR stellar population synthesis} \label{synthesis} \\

\hline \multicolumn{1}{|c|}{\textbf{Galaxy}} & \multicolumn{1}{c|}{\textbf{Ap}} 
     & \multicolumn{1}{c|}{\textbf{x$_y$(\%)}} & \multicolumn{1}{c|}{\textbf{x$_i$(\%)}} 
     & \multicolumn{1}{c|}{\textbf{x$_o$(\%)}}  & \multicolumn{1}{c|}{\textbf{$A_V$}}   
     & \multicolumn{1}{c|}{\textbf{log(t$_{av}$)}}
     & \multicolumn{1}{c|}{\textbf{Z$_{av}$}}  & \multicolumn{1}{c|}{\textbf{M*}[M$_\odot$]}& \multicolumn{1}{c|}{\textbf{M$_{ini}$ }[M$_\odot$]}& \multicolumn{1}{c|}{\textbf{SFR$_{100}$}[M$_\odot$/yr]}
     & \multicolumn{1}{c|}{\textbf{$\chi^2$}}& \multicolumn{1}{c|}{\textbf{adev}}  & \multicolumn{1}{c|}{\textbf{d(")}}\\ \hline 
\endfirsthead

\multicolumn{14}{c}%
{{\bfseries \tablename\ \thetable{} -- continued from previous page}} \\
\hline  \multicolumn{1}{|c|}{\textbf{Galaxy}} & \multicolumn{1}{c|}{\textbf{Ap}} 
     & \multicolumn{1}{c|}{\textbf{x$_y$(\%)}} & \multicolumn{1}{c|}{\textbf{x$_i$(\%)}} 
     & \multicolumn{1}{c|}{\textbf{x$_o$(\%)}}  & \multicolumn{1}{c|}{\textbf{$A_V$}}   
     & \multicolumn{1}{c|}{\textbf{t$_{av}$}}
     & \multicolumn{1}{c|}{\textbf{Z$_{av}$}}  & \multicolumn{1}{c|}{\textbf{M*}[M$_\odot$]}& \multicolumn{1}{c|}{\textbf{M$_{ini}$ }[M$_\odot$]}& \multicolumn{1}{c|}{\textbf{ SFR$_{100}$}[M$_\odot$/yr]}
     & \multicolumn{1}{c|}{\textbf{$\chi^2$}}& \multicolumn{1}{c|}{\textbf{adev}}  & \multicolumn{1}{c|}{\textbf{d(")}}\\ \hline 
\endhead

\hline
\endfoot

\hline \hline
\endlastfoot
NGC0221 & nuc & 0 & 30&  70 & 0.476 &   9.57 &   0.026 &   2.10E+07 & 3.27E+07 & 0.00E+00 & 2.21  &   1.71  &    0.00	  \\*
        & S1  & 0 & 24&  76 &-0.421 &   9.60 &   0.027 &   6.58E+06 & 1.03E+07 & 0.00E+00 & 2.25  &   2.48  &    2.00    \\*
        & S2  & 0 &  5&  95 &-0.428 &   9.76 &   0.023 &   2.32E+06 & 3.70E+06 & 0.00E+00 & 2.31  &   2.24  &    4.00    \\*
        & N1  & 0 & 25&  75 & 0.174 &   9.65 &   0.027 &   8.15E+06 & 1.29E+07 & 0.00E+00 & 2.80  &   1.22  &   -2.00    \\*
\hline
NGC0278 & nuc & 0 & 33&  67 & 0.066 &   9.56 &   0.036 &   7.26E+07 & 1.14E+08 & 0.00E+00 & 1.64  &   2.78  &    0.00    \\*
        & S1  & 0 & 33&  67 &-0.180 &   9.56 &   0.030 &   3.24E+07 & 5.08E+07 & 0.00E+00 & 1.30  &   5.06  &    2.00    \\*
        & S2  & 9 & 25&  65 & 0.210 &   9.56 &   0.032 &   1.50E+07 & 2.33E+07 & 3.37E-03 & 1.17  &  12.31  &    4.00    \\*
        & N1  & 0 & 20&  80 & 0.369 &   9.62 &   0.033 &   3.18E+07 & 5.01E+07 & 0.00E+00 & 1.44  &   7.09  &   -2.00    \\*
\hline
NGC0514 & nuc & 5 & 34&  61 & 0.386 &   9.55 &   0.027 &   1.49E+08 & 2.33E+08 & 2.09E-03 & 0.96  &   5.68  &    0.00    \\*
        & S1  & 0 & 49&  51 & 0.658 &   9.45 &   0.023 &   3.69E+07 & 5.73E+07 & 0.00E+00 & 0.86  &  27.34  &    2.00    \\*
\hline
NGC0674 & nuc &16 & 10&  74 & 1.641 &   9.59 &   0.023 &   1.07E+09 & 1.69E+09 & 4.25E-02 & 1.27  &   4.47  &    0.00    \\*
\hline
NGC0783 & nuc &11 & 48&  42 & 2.001 &   9.41 &   0.024 &   2.05E+09 & 3.14E+09 & 7.39E-02 & 1.13  &   5.29  &    0.00    \\*
        & S1  & 0 &100&	  0 & 1.769 &   8.87 &   0.035 &   3.56E+08 & 4.87E+08 & 0.00E+00 & 0.86  &  11.31  &    2.00    \\*
        & N1  &16 & 64&  19 & 1.922 &   9.17 &   0.033 &   3.21E+08 & 4.69E+08 & 1.74E-01 & 0.85  &  19.00  &   -2.00    \\*
\hline
NGC0864 & nuc &14 & 75&  10 & 1.210 &   8.92 &   0.033 &   3.98E+07 & 5.61E+07 & 1.85E-02 & 0.60  &   6.48  &    0.00	  \\*
        & S1  & 0 & 55&  45 & 0.700 &   9.77 &   0.017 &   3.48E+07 & 5.70E+07 & 0.00E+00 & 0.90  &  18.73  &    2.00    \\*
        & N1  & 0 & 46&  54 & 0.926 &   9.48 &   0.020 &   2.11E+07 & 3.32E+07 & 0.00E+00 & 0.95  &  30.46  &   -2.00    \\*
\hline
NGC1174 & nuc &12 & 88&	  0 & 2.775 &   8.52 &   0.038 &   4.28E+08 & 5.57E+08 & 5.81E-02 & 1.16  &   3.71  &    0.00    \\*
        & S1  & 0 & 42&  58 & 0.357 &   9.57 &   0.026 &   1.64E+08 & 2.55E+08 & 0.00E+00 & 0.68  &  20.17  &    3.00    \\*
        & N1  & 0 & 69&  31 & 1.338 &   9.34 &   0.033 &   1.05E+08 & 1.59E+08 & 0.00E+00 & 0.62  &  45.96  &   -3.00    \\*
\hline
NGC1232 & nuc & 0 &  2&  98  & 0.671 &   9.69 &   0.023 &   1.41E+08 & 2.23E+08 & 0.00E+00 & 0.96  &  18.83  &    0.00    \\
\hline
NGC1482 & nuc &30 & 70&	0 & 3.303 &   8.55 &   0.038 &   1.03E+09 & 1.39E+09 & 4.76E-01 & 1.42  &   3.56  &    0.00    \\
        & S1  & 4 & 59& 37 & 1.518 &   9.71 &   0.039 &   1.46E+09 & 2.39E+09 & 1.28E-02 & 11.05 &  11.05  &   -3.00    \\
\hline
NGC2339 & nuc &48 & 52&	0 & 3.589 &   8.33 &   0.035 &   7.71E+08 & 9.79E+08 & 1.23E+00 & 1.05  &   3.11  &     0.00   \\
        & S1  & 0 & 67&  33 & 2.455 &   9.31 &   0.037 &   4.03E+08 & 6.08E+08 & 0.00E+00 & 0.81  &   7.82  &     2.40   \\
        & N1  &42 & 58&	0 & 3.218 &   8.51 &   0.035 &   1.13E+08 & 1.48E+08 & 1.40E-01 & 0.83  &  10.38  &    -2.40   \\
\hline
NGC2342 & nuc &26 & 74&	0 & 2.432 &   8.56 &   0.036 &   1.28E+09 & 1.68E+09 & 4.18E-01 & 1.14  &   4.39  &     0.00   \\
\hline
NGC2903 & nuc & 0 &  0& 100 & 1.598 &   9.70 &   0.032 &   2.30E+08 & 3.63E+08 & 1.67E-08 & 2.23  &   2.72  &     0.00   \\
        & S1  & 6 &  0&  94 & 1.361 &   9.67 &   0.037 &   9.91E+07 & 1.57E+08 & 1.25E-03 & 1.95  &   3.27  &     1.66   \\
        & S2  &30 &  3&  67 & 0.805 &   9.53 &   0.032 &   8.99E+07 & 1.42E+08 & 7.98E-03 & 2.18  &   2.97  &     3.58   \\
        & S3  &19 &  0&  81 & 0.551 &   9.61 &   0.034 &   3.38E+07 & 5.36E+07 & 1.64E-03 & 1.47  &   4.68  &     4.98   \\
        & N1  & 2 &  3&  95 & 1.090 &   9.68 &   0.037 &   9.32E+07 & 1.48E+08 & 4.15E-04 & 2.08  &   3.67  &    -1.74   \\
        & N2  &56 &  8&  36 & 0.895 &   9.27 &   0.023 &   4.34E+07 & 6.78E+07 & 1.27E-02 & 1.90  &   3.16  &    -2.40   \\
        & N3  &45 & 55&	0 & 1.873 &   8.03 &   0.036 &   1.22E+07 & 1.53E+07 & 9.41E-03 & 1.38  &   4.13  &    -4.63   \\
        & N4  &45 &  1&  54 & 1.357 &   9.43 &   0.040 &   2.72E+07 & 4.29E+07 & 4.59E-03 & 1.34  &   5.67  &    -6.13   \\
\hline
NGC2950 & nuc & 0 & 23&  77 & 0.579 &   9.60 &   0.031 &   2.22E+09 & 3.49E+09 & 0.00E+00 & 2.41  &   2.27  &     0.00   \\
        & S1  & 0 & 21&  79 &-0.427 &   9.61 &   0.027 &   3.89E+08 & 6.11E+08 & 0.00E+00 & 2.16  &   3.43  &     1.50   \\
        & S2  & 0 & 22&  78 &-0.510 &   9.61 &   0.026 &   1.66E+08 & 2.59E+08 & 0.00E+00 & 1.14  &   5.06  &     2.50   \\
        & N1  & 0 & 24&  76 & 0.500 &   9.60 &   0.028 &   4.45E+08 & 6.97E+08 & 0.00E+00 & 2.23  &   3.26  &    -1.50  \\ 
        & N2  & 0 & 18&  82 & 0.329 &   9.62 &   0.025 &   1.94E+08 & 3.04E+08 & 0.00E+00 & 1.27  &   5.49  &    -2.50   \\
\hline
NGC2964 & nuc &12 & 88&	0 & 2.155 &   8.39 &   0.036 &   3.23E+08 & 4.15E+08 & 4.65E-02 & 0.97  &   5.41  &     4.98   \\
        & S1  & 8 & 92&	0 & 1.730 &   8.28 &   0.038 &   8.66E+07 & 1.08E+08 & 5.91E-02 & 1.08  &  21.17  &     0.00   \\
\hline
NGC3184 & nuc &21 & 79&	0 & 1.471 &   8.27 &   0.039 &   8.97E+06 & 1.13E+07 & 1.41E-02 & 0.85  &   7.01  &     0.00   \\
        & S1  &18 & 82&	0 & 1.155 &   7.93 &   0.040 &   3.22E+06 & 3.99E+06 & 1.22E-03 & 0.74  &  18.95  &     2.00   \\
        & N1  & 3 & 97&	0 & 1.675 &   7.99 &   0.040 &   4.06E+06 & 5.05E+06 & 1.07E-04 & 0.75  &  24.56  &    -2.00   \\
\hline
NGC4102 & nuc &52 & 48&	0 & 3.251 &   8.58 &   0.020 &   8.02E+08 & 1.05E+09 & 2.02E+00 & 1.00  &   2.72  &      0.00  \\
        & S1  &58 & 42&	0 & 2.528 &   8.42 &   0.018 &   9.61E+07 & 1.29E+08 & 1.33E-01 & 1.04  &   3.58  &      2.00  \\
        & N1  &38 & 62&	0 & 3.446 &   8.54 &   0.032 &   3.14E+08 & 3.97E+08 & 8.26E-01 & 1.37  &   3.30  &     -2.00  \\
\hline
NGC4179 & nuc & 0 &  4&96 & 0.951 &   9.69 &   0.026 &   2.17E+09 & 3.42E+09 & 0.00E+00 & 1.96  &   4.24  &      0.00  \\
        & S1  & 0 &  2&98 & 0.587 &   9.69 &   0.024 &   7.18E+08 & 1.13E+09 & 0.00E+00 & 1.45  &   6.41  &      2.00  \\
\hline
NGC4303 & nuc & 0 & 55&45 & 0.557 &   9.48 &   0.033 &   3.12E+08 & 4.81E+08 & 0.00E+00 & 0.87  &   2.98  &      0.00  \\*
        & S1  & 0 & 31&69 & 0.357 &   9.56 &   0.034 &   7.76E+07 & 1.22E+08 & 0.00E+00 & 1.77  &   3.45  &      1.50  \\*
        & S2  &34 & 30&36 & 0.575 &   9.30 &   0.039 &   2.91E+07 & 4.45E+07 & 2.69E-02 & 1.14  &   4.40  &      2.50  \\*
        & S3  &17 & 24&59 & 0.703 &   9.49 &   0.035 &   2.14E+07 & 3.36E+07 & 1.15E-03 & 0.70  &   6.29  &      3.50  \\*
        & N1  & 0 & 33&66 & 0.431 &   9.55 &   0.035 &   7.09E+07 & 1.11E+08 & 1.52E-05 & 1.60  &   3.63  &     -1.50  \\*
        & N2  &23 & 42&35 & 1.085 &   9.30 &   0.036 &   2.76E+07 & 4.21E+07 & 1.49E-02 & 1.05  &   4.87  &     -2.50  \\*
        & N3  &33 & 25&42 & 1.017 &   9.36 &   0.035 &   1.47E+07 & 2.27E+07 & 7.81E-03 & 0.81  &   9.51  &     -3.50  \\*
\hline
NGC4461 & nuc &48 & 28&  24 & 1.875 &   9.51 &   0.014 &   7.15E+08 & 1.15E+09 & 7.55E-01 & 0.53  &   4.91  &      0.00  \\
        & S1  &24 & 60&  16 & 1.609 &   9.36 &   0.033 &   3.18E+08 & 4.82E+08 & 3.05E-01 & 0.94  &   8.11  &      2.00  \\
        & N1  & 0 & 79&  21 & 0.546 &   9.22 &   0.030 &   1.80E+08 & 2.69E+08 & 0.00E+00 & 0.95  &   9.06  &     -2.00  \\
\hline
NGC4845 & nuc &35 & 65&	1 & 4.000 &   8.55 &   0.029 &   5.96E+08 & 7.93E+08 & 4.63E-01 & 1.92  &   2.41  &      0.00  \\
        & S1  & 9 & 91&	0 & 2.555 &   8.41 &   0.033 &   1.29E+08 & 1.67E+08 & 1.22E-02 & 1.24  &   3.60  &      1.50  \\
        & S2  &25 & 75&	0 & 3.039 &   8.37 &   0.024 &   6.54E+07 & 8.16E+07 & 1.19E-01 & 0.97  &   6.52  &      2.50  \\
        & N1  &51 & 49&	0 & 4.000 &   8.43 &   0.032 &   8.87E+07 & 1.12E+08 & 2.67E-01 & 1.08  &   4.42  &     -1.50  \\
        & N2  &50 & 50&	0 & 4.000 &   8.54 &   0.030 &   3.92E+07 & 4.90E+07 & 1.40E-01 & 0.94  &   8.52  &     -2.50  \\
\hline
NGC5457 & nuc &64 & 28&	7 & 1.182 &   8.74 &   0.016 &   1.13E+07 & 1.60E+07 & 3.82E-02 & 0.72  &   5.83  &      0.00  \\
        & S1  &54 & 46&	0 & 1.925 &   8.13 &   0.031 &   5.13E+06 & 5.97E+06 & 2.36E-02 & 0.84  &  15.96  &      2.00  \\
        & N1  &41 & 59&	0 & 2.074 &   8.29 &   0.033 &   4.57E+06 & 5.82E+06 & 9.48E-03 & 0.87  &  20.81  &     -2.00  \\
\hline
NGC5905 & nuc &32 & 20&  49 & 1.393 &   9.41 &   0.023 &   1.35E+09 & 2.12E+09 & 1.61E-01 & 1.32  &   3.31  &      0.00  \\
        & S1  &52 & 23&  25 & 1.537 &   9.15 &   0.038 &   2.96E+08 & 4.42E+08 & 5.77E-01 & 0.87  &   6.97  &      2.00  \\
        & N1  &46 & 42&  12 & 1.633 &   9.06 &   0.031 &   2.73E+08 & 3.97E+08 & 4.40E-01 & 0.92  &   7.57  &     -2.00  \\
\hline
NGC6181 & nuc &38 & 33&  30 & 1.195 &   9.27 &   0.030 &   6.95E+08 & 1.06E+09 & 3.40E-01 & 1.10  &   3.50  &      0.00  \\
        & S1  & 5 & 27&  67 & 0.292 &   9.59 &   0.034 &   4.72E+08 & 7.39E+08 & 6.38E-03 & 0.93  &   5.07  &      2.00  \\
\hline
NGC6946 & nuc &56 & 44&	0& 3.295 &   8.41 &   0.033 &   6.29E+07 & 8.29E+07 & 9.61E-02 & 2.83  &   1.33  &      0.00  \\
\hline
NGC7080 & nuc &16 & 67&  17 & 1.725 &   9.17 &   0.029 &   2.32E+09 & 3.44E+09 & 2.95E-01 & 1.28  &   2.54  &  0.00 	  \\
        & S1  & 0 & 31&  69 & 0.890 &   9.59 &   0.036 &   1.22E+09 & 1.91E+09 & 1.49E-06 & 1.01  &   5.83  &      2.00  \\
        & N1  & 0 & 56&  44 & 1.358 &   9.39 &   0.036 &   7.80E+08 & 1.20E+09 & 0.00E+00 & 0.98  &   7.09  &     -2.00  \\
\hline
NGC7448 & nuc & 3 & 43&  55 & 0.196 &   9.44 &   0.036 &   1.66E+08 & 2.56E+08 & 1.45E-03 & 0.92  &   5.41  &      0.00  \\
        & S1  & 0 & 37&  63 & 0.298 &   9.50 &   0.036 &   1.06E+08 & 1.64E+08 & 0.00E+00 & 0.75  &   8.84  &      2.00  \\
        & N1  & 0 & 29&  71 & 0.085 &   9.55 &   0.033 &   1.06E+08 & 1.66E+08 & 0.00E+00 & 0.83  &  10.22  &     -2.00  \\
\hline
NGC7798 & nuc &10 & 25&  65 & 1.473 &   9.53 &   0.025 &   1.20E+09 & 1.88E+09 & 3.53E-02 & 1.47  &   3.55  &      0.00  \\
        & S1  & 0 & 10&  90 & 1.195 &   9.75 &   0.026 &   2.96E+08 & 4.77E+08 & 0.00E+00 & 0.70  &  10.28  &      2.00  \\
        & N1  & 0 & 37&  63 & 1.828 &   9.58 &   0.011 &   2.22E+08 & 3.51E+08 & 0.00E+00 & 0.66  &  22.11  &     -2.00  \\
\hline
NGC7817 & nuc & 0 & 15&  85 & 2.041 &   9.65 &   0.031 &   5.78E+08 & 9.09E+08 & 0.00E+00 & 1.03  &   5.35  &      0.00  \\
        & N1  & 0 &  1&  99 & 1.538 &   9.70 &   0.022 &   1.84E+08 & 2.90E+08 & 0.00E+00 & 0.80  &  18.17  &     -2.00  \\

\hline

\label{specfitresult}
\end{longtable}
\end{center}

\twocolumn

\subsection{Comparison with higher resolution models}

Because higher resolution stellar population models are available in the literature for
the wavelength range covered by our observations, it is important to test 
these models and compare the output with that of the lower resolution. 
In principle, one would expect that higher resolution models produce a higher 
accuracy in the output since the signatures we are looking for should be more conspicuous.
and clear. The main limitation, however, is that \citet{maraston+11} models are available 
only for solar metallicity.
They were constructed using the same recipe as the ones from
\citet{maraston05}, but using different observables. The set of 
models that extends to the NIR are the ones that use the
\citet{pickles98}'s stellar spectral library and above $\sim$ 1~$\mu$m, nearly half
of the spectra from that library lack of spectroscopic observations.
In order to solve this problem, the author constructed a smooth energy distribution from
broadband photometry. It may imply that some NIR absorption
features may not be well resolved, even for these higher resolution
models. 

\begin{figure*} 
\includegraphics [width=162mm]{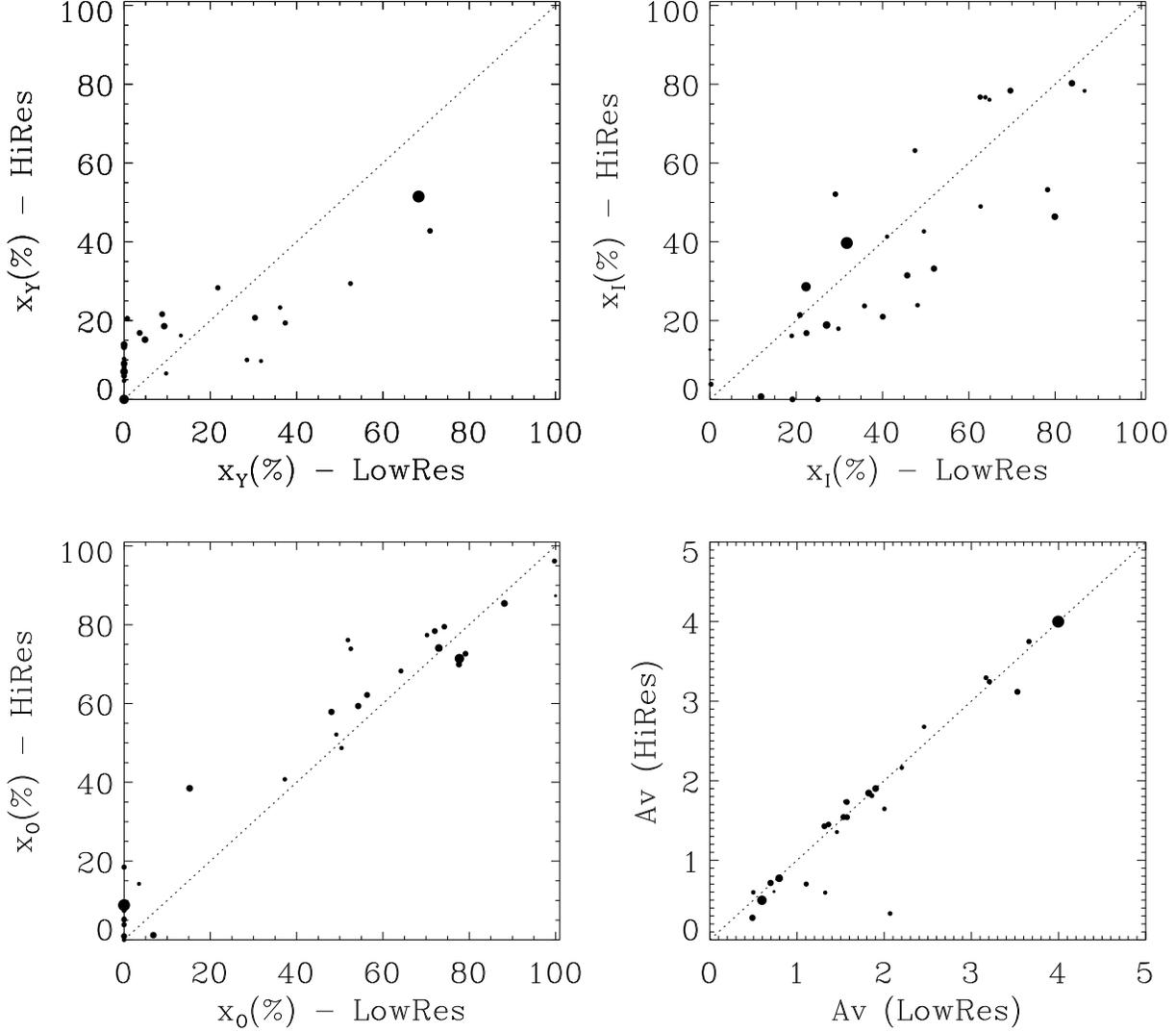}
\caption{Comparison between the results of the spectral synthesis done with low resolution models
\citep{maraston05} and with the high resolution ones \citep{maraston+11}. Only solar metallicity
was used for this test. The comparison is done for x$_Y$ (top left), x$_I$ (top right), x$_O$ (bottom left)
and A$_V$. The dotted line shows the one to one relation. The size of the points is inversely proportional
to the {\it adev} value found by the fitting procedure.}
\label{comp_lowhires}
\end{figure*}

We applied the synthesis population fitting method to our galaxies 
using these high resolution SSPs for the base, 
using the same ages as described in section 3.
These results were compared with the low resolution models. However, 
to be certain that differences were due only to model differences and
not metallicity differences, we 
performed
the stellar population synthesis with 
the low resolution models again,
but with a base of SSP containing only solar metallicity models. This
will ensure that any differences found between the results were not due
to any age-metallicity degeneracy effect. 
The comparison between the results
of the low and high resolution SSPs can be seen in Figure~\ref{comp_lowhires}. 
Note that only nuclear apertures are shown in the plot because of their
higher S/N, reducing uncertainties in the fit. The dotted
line in Figure~\ref{comp_lowhires} marks the one to one relationship. The size
of each point in these plots is inversely proportional to the
$adev$ value found by the fitting procedure, meaning that
the larger the point, the smaller the error.

It can be seen from Figure~\ref{comp_lowhires} that the fraction 
of old population found for each galaxy does not
change significantly when using the high resolution models instead of the
low resolution ones (bottom left panel). The same can be said
about the reddening (bottom right panel). Regarding the fraction of 
intermediate age stellar populations (top right panel), discrepancies are 
clearly observed, but still most points are distributed 
along the one-to-one relationship. The test fails, as expected, for the 
fraction of young stellar population (top left panel).  

Indeed, \citet{maraston+11} found that the larger discrepancies between
the old models and the new ones are in the NIR, for ages around 10 $-$ 20 Myr,
where the red supergiants strongly contributes. Due to the short duration
of their phase, a large scatter is expected for models
that try to predict their contribution to the continuum. 
Besides, since the higher resolution models do not
have the features expected for this wavelength region,
they tend to be bluer than the low resolution models. 
This would explain why the synthesis with these models find a smaller
contribution of x$_Y$ to the spectra.

Given the limitations of the spectral library used for the high
resolution models in the NIR, it is not clear that they bring
an improvement to the method that compensates the fact that they
are available only for solar metallicity. For this reason,
we consider that the low resolution models of \citet{maraston05} are still 
the best option of SSPs for this wavelength range.

\subsection{Spectral Synthesis and the NIR Indexes}

Since the development of stellar population models that take into account 
more rigorously the TP-AGB phase, the CN molecular band has been
advertised as an unambiguous evidence of intermediate-age stellar population
\citep{maraston05, riffel+07, riffel+09}. This issue, however, has not yet
been fully assessed on observational grounds, mostly because the results 
gathered up to now are based on samples containing only a few star-forming 
galaxies or active galactic nuclei.

\citet{riffel+09}, for example, presented a histogram comparing the 
fraction of intermediate age stellar population of 
Seyfert galaxies (obtained by stellar population synthesis)
with the presence or absence of the CN band. They found
a slight tendency of the galaxies that display CN to
show a higher fraction of intermediate-age stellar population.
Note, however, that they also report galaxies with no CN and a high 
fractions of x$_I$ as well as galaxies with CN and low fractions 
of x$_I$. Later, \citet{zibetti+12} claimed that the two main signatures
of the presence of TP-AGB stars predicted by the \citet{maraston05}'s models,
CN at 1.41$\mu$m and C$_2$ at 1.77$\mu$m, are not detected in a 
sample of 16 post-starburst galaxies.

\begin{figure}
\includegraphics [width=86mm]{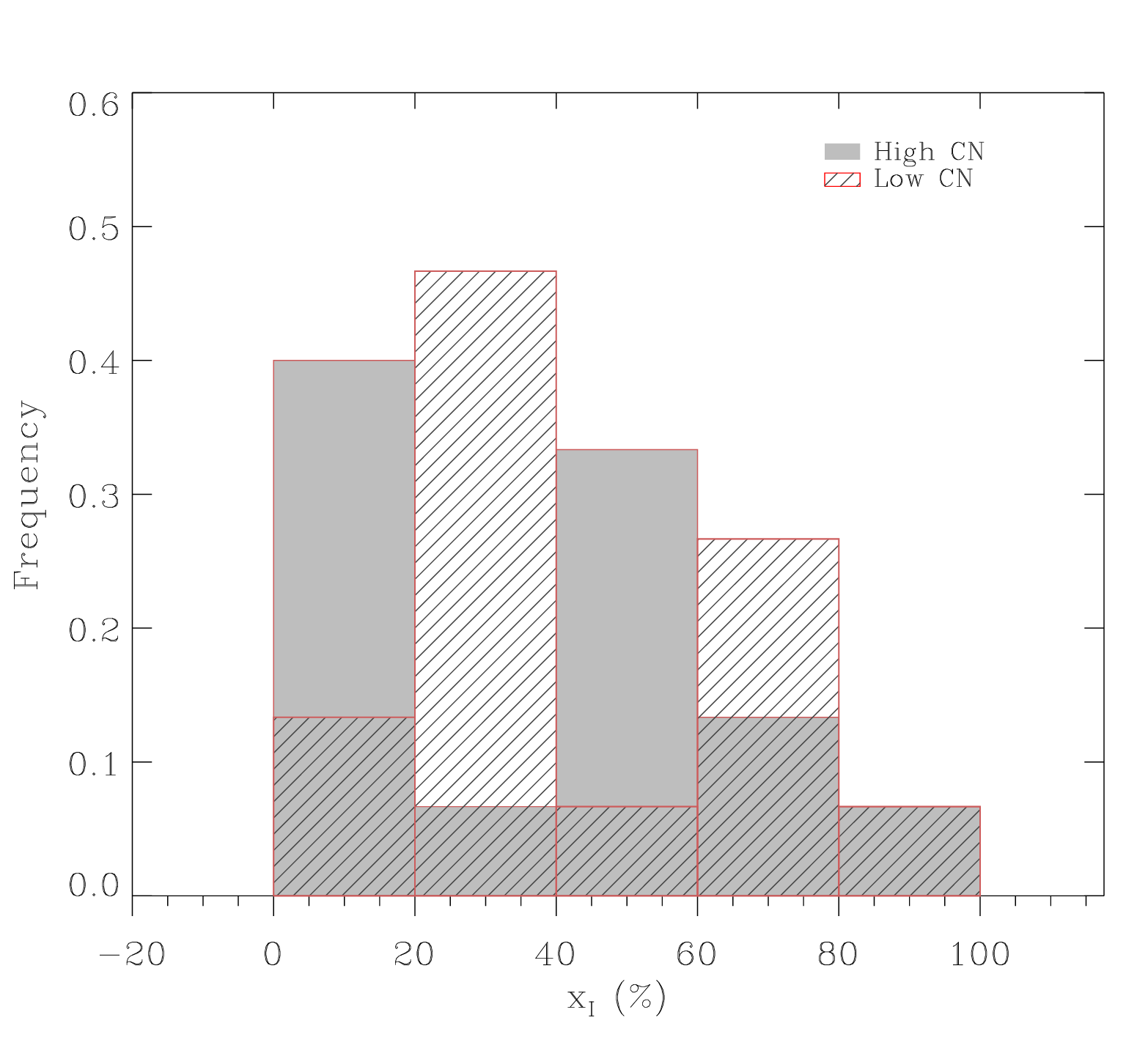}
\caption{Histogram comparing the intermediate age component (x$_I$) for the nuclear
aperture of the galaxies with low values of CN (crosshatch) and high
values of CN (shaded in gray). }
\label{CN_comp}
\end{figure}

Figure~\ref{CN_comp} shows the relation between the presence of the CN band 
and the intermediate age population for our galaxies. Only 
nuclear apertures were considered. The values 
of the CN index (equivalent width as defined by \citet{riffel+07}) were obtained from \citet{martins+13}, with an average value
of 18.3 \AA. We divided our sample into two groups: galaxies with low values 
of CN (values below the average) and high values of CN (values above the average). 

From the histogram, there is no clear trend between the CN index
and the fraction of intermediate stellar population within a galaxy.
We found galaxies with a very high fraction of intermediate-age
stellar population with no measured CN and galaxies where the CN is very
strong but only a small fraction of intermediate-age stars was found.
As pointed out by \citet{riffel+09}, that result does not necessarily imply 
that the CN is not a suitable tracer of intermediate-age stellar population. 
The proximity of strong permitted lines to the CN band may hinder, for instance, 
its observation. This is the case of galaxies where He\,{\sc i}~1.081\,$\mu$m 
and Pa$\gamma$ in emission are strong. The red wings of these two lines will
partially or totally fill up the CN band. As a result, the derived 
CN index will be small. This problem is particularly relevant in
AGNs with broad permitted lines, which is not our case. Another possibility is
that for galaxies with low $z$ ($z<$0.01), the CN band falls in a region of 
strong telluric absorption. Residuals left after the division by the 
telluric standard star may affect the S/N around 1.1~$\mu$m. In this 
case, the measurement of the CN band may be subjected to large 
uncertainties.

For our sample, we estimate that low S/N around 1.1~$\mu$m can be 
an issue in NGC\,1232 (x$_{\rm i}$=97.5\%), NGC\,2964 (x$_{\rm i}$=87.5\%), 
NGC\,3184 (x$_{\rm i}$=78.8\%), NGC\,4461 (x$_{\rm i}$=27.8\%) and 
NGC\,7817 (x$_{\rm i}$=15\%).  Note, however, that NGC\,2964 and NGC\,3184 
display emission lines in their NIR spectra, giving additional support to 
the result of a large fraction of intermediate-age stars in their nuclei
(see next section). 

We found that although the CN band is a potential tracer of the 
intermediate age population, in practice its usefulness can be 
limited by the proximity of emission lines and residuals left after 
the division by the telluric standard.

\subsection{Stellar population synthesis results and the presence of emission lines}

Although no correlation was found between the stellar population synthesis  
and the NIR indexes, we were able to identify a 
relationship between the stellar population and the presence of emission lines 
in the galaxy sample. 
Galaxies that display no emission lines in the NIR were found to have, in general,
very old stellar populations. Assuming that the spectra recorded for these sources is representative of 
the bulk of their stellar populations, these objects are either dominated by x$_O$ or 
it represents a large fraction of the emitted light. The weak nebular component found
in the optical region contributes little to the NIR spectrum and/or the regions that
emit these lines were not present inside the slit. Most of the important young
population contributions were found in apertures with strong emission lines. Only a few
were found in off-nuclear apertures with no signs of emission lines, but this
happens when the S/N of the spectrum is very low, which means larger uncertainties
in the fit. This is observed by the high {\it adev} value of these fits.

\citet{martins+13} compared emission line properties from the optical
observations and from the NIR. They found that for a subsample of galaxies the emission
lines in the NIR were much weaker than in the optical, sometimes even absent.
They concluded that this was mainly due to the differences in aperture between
these two sets of data - the aperture in the optical has 5 times the area than
the aperture in the NIR. In many of these galaxies the star-formation is probably
not nuclear, but circumnuclear, or located in hot spots outside the nucleus. 
Based on this comparisons, they classified the galaxy sample in four classes:

\begin{itemize}

\item Weak emission lines in the optical, no emission lines in the NIR,
either at the nucleus or
in the extended region (class 1): NGC\,514, NGC\,674, NGC\,6181, NGC\,7448.

\item Strong emission lines in the optical, no emission lines in the NIR, 
either at the nucleus or
in the extended region (class 2): NGC\,278, NGC\,7817

\item Strong emission lines in the optical, evidence of weak to moderate-intensity lines 
in the NIR (nucleus or/and extended region - class 3):  
NGC\,783, NGC\,864, NGC\,1174, NGC\,2964, NGC\,3184, NGC\,4303, NGC\,4845, NGC\,5457, NGC\,5905, NGC\,7080,
NGC\,7798


\item Strong emission lines in the optical, moderate to strong emission lines in the NIR 
(class 4):  
NGC\,2339, NGC\,2342, NGC\,2903, NGC\,4102, NGC\,6946

\end{itemize}

\begin{figure*}
\includegraphics [width=178mm]{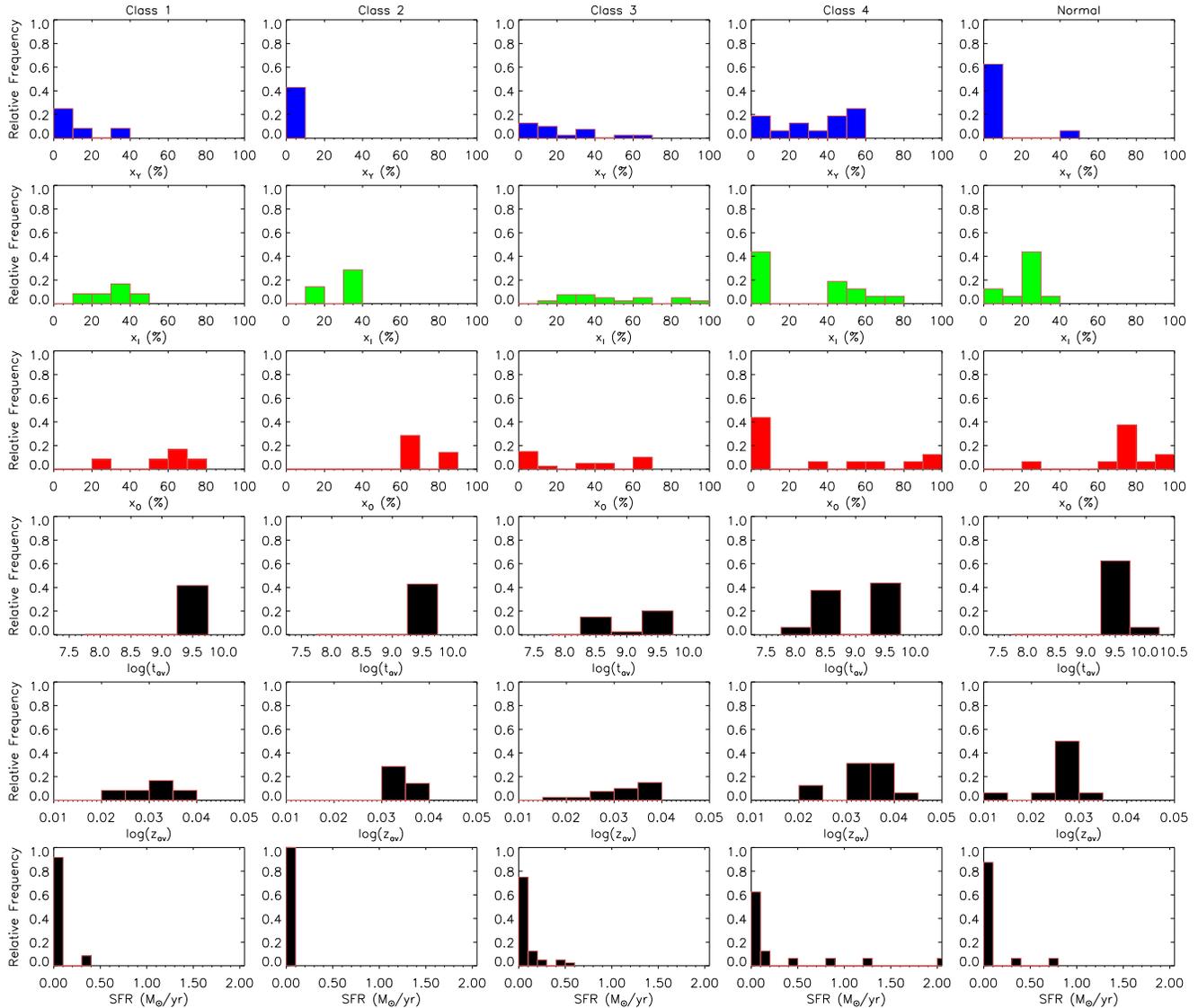}
\caption{Histogram comparing the results of the synthesis separated by the classes defined
in \citet{martins+13} }
\label{synperclass}
\end{figure*}

Figure~\ref{synperclass} shows the synthesis results for each of these
classes. Only results for which the {\it adev} value was smaller
than 5.0 were included in these plots. This figure shows the 
distribution of x$_Y$, x$_I$, x$_O$, the average age (t$_{av}$)
and average metallicity (Z$_{av}$) of the stellar population
and the star formation rate  (SFR$_{100}$) in the last 100 Myrs, as 
found by the synthesis.

All class~1 and class~2 objects have very old average
age. For all objects the synthesis found a very old stellar 
population in the nucleus. The lack of emission lines
in the NIR was interpreted by \citet{martins+13} as a difference
between the slit sizes of the optical observations and the ones
done in the NIR. Since the slit used for the NIR observations was much smaller, 
the ionisation source was missed by these observations, and that 
agrees with the old ages found by the synthesis.
NGC~6181 seems to be 
a particular exception since a contribution of x$_Y$ to the nucleus larger than 30$\%$ was found. 
However, its average age is still high.
For the off-nuclear apertures of NGC~514 and NGC~674 the spectra were very noisy and the
synthesis results cannot be really trusted.

Class~3 and 4 objects have in general much lower average ages. All class~4
and most of class 3 objects have significant contribution of a young stellar population.
It is also very interesting that this contribution tends to be
higher in the apertures where the emission lines are stronger. 
This young stellar population should be the responsible
for the gas ionisation. We also found higher star formation
rates in the last 100 Myrs in objects from classes 3 and 4. There
are no significant differences in metallicity between the classes. However,
the non-star-forming galaxies seem to have lower metallicities than the galaxies
classified as starburst or star-forming.

It is also interesting to comment on the results found for the non-star-forming galaxies NGC~221, 
NGC~1232, NGC~2950, NGC~4179 and NGC~4461. 
The first four galaxies, as expected, are dominated by x$_O$. The lack of young stars able
to ionise the gas should explain the absence of emission lines in their spectra. 
However NGC~4461 has an important contribution of x$_Y$. The result for this galaxy is 
somewhat puzzling because the quality of the fit, as measured by the {\it adev} value
is not bad (at least for the nuclear aperture). 
We do not expect this galaxy to have any contribution of a young, ionising 
stellar population to its continuum, otherwise strong emission lines should be seen.
One possibility is that this young population is indeed real and the galaxy has no gas
(or not enough to have strong emission lines). 

Taking all this together, we found that overall the stellar population synthesis
is recovering what would be expected of star-forming galaxies. However, individual results 
have to be carefully verified.

\subsection{Stellar Extinction}

\begin{figure*}
\includegraphics [width=182mm]{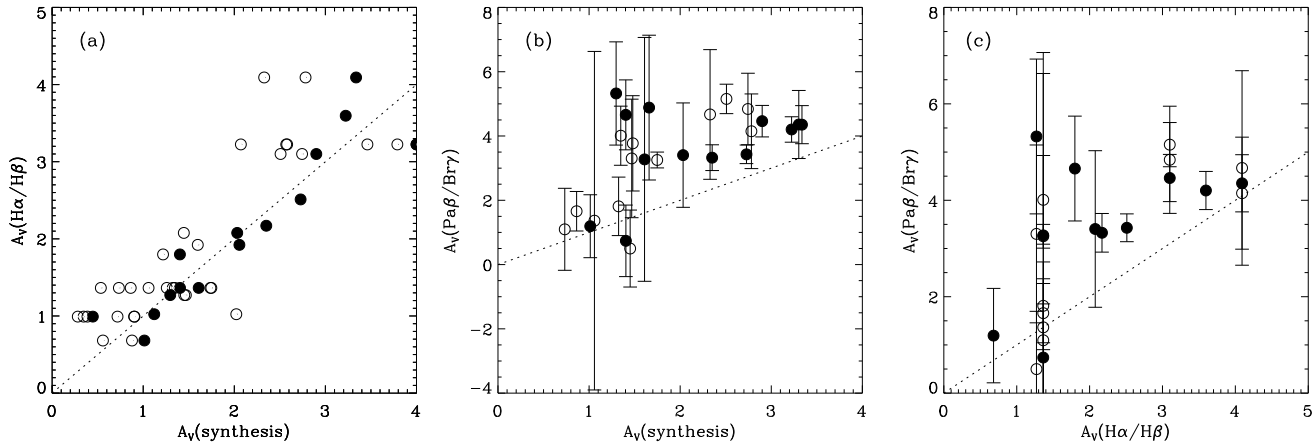}
\caption{Comparison between the A$_{V}$ extinction values measured from the stellar 
population synthesis in the NIR (synthesis), from the H$\alpha$/H$\beta$ line ratio from HO97 
and from the Pa$\beta$/Br$\gamma$ measured in the NIR spectra. Filled circles 
represent the nuclear apertures and open circles represent the off-nuclear apertures.}
\label{extinction}
\end{figure*}

\citet{martins+13} calculated the gas extinction in the NIR
using the hydrogen ratio Pa$\beta$/Br$\gamma$, and found it
to be generally larger than the gas extinction calculated from 
optical emission lines. This result agreed with previous results 
from the literature, and is explained by the fact that the NIR
should probe larger optical depths than the optical range. 
One of the outputs from the stellar populations synthesis is
A$_V$, the extinction from the continuum. Figure~\ref{extinction}
shows the comparison of the values we obtained with the gas extinction 
from \citet{martins+13}. Filled circles represent the nuclear and
open circles represent the off-nuclear apertures.

The extinction we found from the stars in the NIR traces the extinction
from the gas in the optical. This is somewhat expected. The extinction
obtained from the gas in the optical is higher than the one derived 
from the stars in this same wavelength region because the
gas emission comes from a dustier region than the stars. However, the 
NIR emission from the stars should also probe dustier regions than the optical emission
from the stars, which means that the emission from the gas in the optical and the
continuum emission from the stars in the NIR should 
probe similar optical depths.
However, one has to keep in mind that they do not have to be associated 
with the same spatial positions, as they differ from the extinction 
from the stars in the optical for different reasons - the gas
emission intrinsically arrises from dustier positions while the
stars in the NIR just probe higher optical depths.

\begin{figure}
\includegraphics [width=86mm]{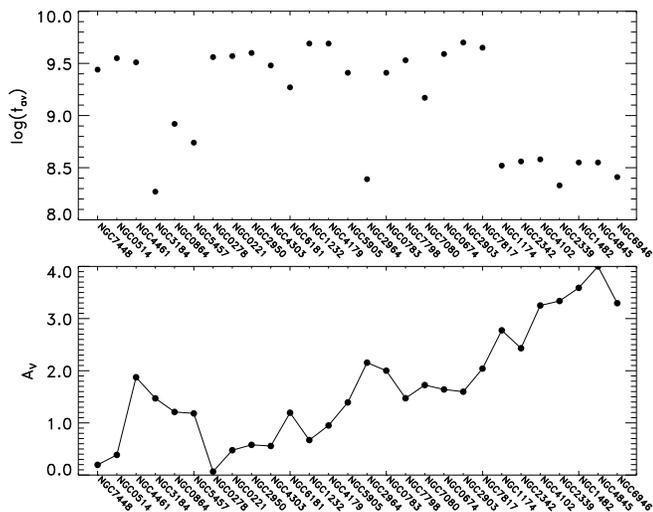}
\caption{Relation between the average age (top panel) and the continuum extinction in the NIR (bottom
panel) with the continuum steepness. The abscissa shows the galaxies from the sample in order
of continuum steepness, from steepest ones (left) to the flatter ones (right).}
\label{ordercont}
\end{figure}

\citet{martins+13} also showed that the continuum of the galaxies in the
NIR have a diversity of shapes and steepness. In Figure~6 of their paper,
they show the nuclear spectra of the galaxies, organised by
their steepness. They suggest that the
shapes are related to the presence or absence of dust and the young
stellar population. Here we present in Figure~\ref{ordercont} 
the average age and the stellar reddening obtained by the synthesis
for the galaxies, now organised in steepness order, from the steepest continua (left)
to flatter ones (right).

From this figure it is clear that the continuum shape is closely related to the 
stellar extinction (bottom panel), and that the presence of younger
stellar populations make the continuum flatter.

\subsection{Comparison with Far-Infrared Tracers }

Galaxies with active star-formation present a peak of emission around 60 $\mu$m,
which is the maximum temperature of the dust heated by this process. 
The young OB stars that dominate the starburst radiate primarily in the optical and ultraviolet, but surrounding gas and dust reprocesses this radiation and thus strongly radiates at thermal wavelengths in the far-infrared. Far-IR luminosity (LIR) is thus indicative of the magnitude of recent star formation activity \citep{telesco88,lonsdale+84}. 

We can then try to correlate the LIR with our results. Again, one has to be careful 
with the aperture sizes in each of the observations, to be sure results are consistent.
We obtained IRAS (Infrared Astronomical Satellite) IR fluxes from \citet{surace+04}, 
and calculated the LIR luminosity as defined in \citet{sanders+96}. 

\begin{figure*}
\includegraphics [width=172mm]{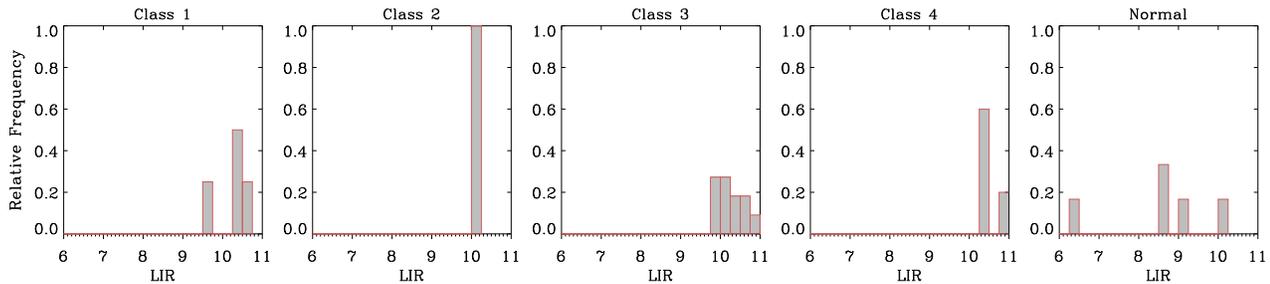}
\caption{Distribution of LIR luminosities for the emission line classes defined by \citet{martins+13}}
\label{lirdistrib}
\end{figure*}

As expected, no direct correlation was found between these values and the fractions
of x$_Y$ or x$_I$ found by the synthesis. This is directly related to the fact that
for many of the galaxies in our sample the star-formation is not a nuclear phenomenon. 
Figure~\ref{lirdistrib} shows the LIR distribution separated by the classes as defined 
by \citet{martins+13}. From this figure it is clear that classes 3 and 4, which had the strongest
emission lines and higher fractions of younger populations, present also higher LIR values in
comparison with classes 1 and 2. Non-star-forming galaxies clearly have the lowers LIR luminosities, as
expected. 

\section{Summary and conclusions}

We performed stellar population synthesis in a sample of
long-slit NIR spectra of 23 star-forming and 5 non-star-forming galaxies
to test  the predictions of stellar population models that are available in this wavelength region.
The stellar population synthesis code used for this work was 
STARLIGHT. The chosen SSP models used as a base for STARLIGHT were the ones from \citet{maraston05},
because they include the effect of the TP-AGB phase, crucial to model the stellar population
in this wavelength region. So far, these (and their most recent version in \citet{maraston+11}) are the only
ones published that take this effect into account. 

We compared the synthesis results from older version of the models \citep{maraston05} with the new version 
\citet{maraston+11}. 
We found discrepancies for the fraction of young stellar populations, which were mentioned 
by the own authors, due to the uncertainties in the presence of the red supergiants. 
Given the limitations of these models, we believe that the lower resolution models from \citet{maraston05}
are still the best set of models for stellar population synthesis in the NIR.

No hot dust contribution was found for any galaxy, ruling out the presence of a possible
hidden AGN in any of them.
We found no correlation between the synthesis results and the NIR indexes measured by
\citet{martins+13}. Although some of the signatures, like the CN band, 
can be potential traces of the intermediate age population, we believe that 
the use of them in practice is still complicated, mostly because of observational limitations.

The results from the synthesis seem to be in very good agreement with the emission line
measurements. For galaxies with no emission lines detected in the NIR, no significant
young stellar population was found. They have their continuum dominated by old stellar populations.
In addition, the apertures with strong emission lines have large contribution of a young stellar population.
This becomes more obvious when comparing the results from the synthesis separated by the classes 
defined by \citet{martins+13}. The classes were created based on the strength of the emission
lines in the NIR, with class 1 objects presenting no detected emission lines in the NIR and
class 4 having the strongest emission lines measured. Classes 1 and 2 objects have higher contribution
of older stellar populations and higher average ages. Classes 3 and 4 objects have
higher contribution of the young stellar population and lower average ages.

From the synthesis we also obtained the extinction from the continuum in the NIR, which we
found to trace the extinction from the gas in the optical. We also find that the 
continuum steepness is related to this extinction, in the sense that the steepest continua
tend to have smaller extinction values that the flatter ones.

\section*{Acknowledgments}
The authors thank the reviewer of the paper for his valuable comments.
This research has been partially supported by the Brazilian agency FAPESP (2011/00171-4).
ARA thanks to CNPq for financial support through grant 307403/2012-2. 
RR thanks to FAPERGs (ARD 11/1758-5) and
CNPq (304796/2011-5) for financial support.


\label{lastpage}


\begin{thebibliography}{}
\expandafter\ifx\csname natexlab\endcsname\relax\def\natexlab#1{#1}\fi

\bibitem[{Alonso-Herrero {et~al}\mbox{.}(2000)Alonso-Herrero, Rieke, Rieke, \&
  Shields}]{alonso-herrero+00}
Alonso-Herrero A., Rieke M.~J., Rieke G.~H., Shields J.~C., 2000, The
  Astrophysical Journal, 530, 688

\bibitem[{Asari {et~al}\mbox{.}(2007)Asari, {Cid Fernandes}, Stasi\'nska,
  Torres-Papaqui, Mateus, Sodr\'{e}, Schoenell, \& Gomes}]{asari+07}
Asari N.~V., {Cid Fernandes} R., Stasi\'nska G., Torres-Papaqui J.~P., Mateus
  A., Sodr\'{e} L., Schoenell W., Gomes J.~M., 2007, Monthly Notices of the
  Royal Astronomical Society, 381, 263

\bibitem[{Balogh {et~al}\mbox{.}(1997)Balogh, Morris, Yee, Carlberg, \&
  Ellingson}]{balogh+97}
Balogh M.~L., Morris S.~L., Yee H. K.~C., Carlberg R.~G., Ellingson E., 1997,
  The Astrophysical Journal, 488, L75

\bibitem[{{Burston} {et~al}\mbox{.}(2001){Burston}, {Ward}, \&
  {Davies}}]{burston+01}
{Burston} A.~J., {Ward} M.~J., {Davies} R.~I., 2001, \mnras, 326, 403

\bibitem[{{Calzetti} {et~al}\mbox{.}(2000){Calzetti}, {Armus}, {Bohlin},
  {Kinney}, {Koornneef}, \& {Storchi-Bergmann}}]{calzetti+00}
{Calzetti} D., {Armus} L., {Bohlin} R.~C., {Kinney} A.~L., {Koornneef} J.,
  {Storchi-Bergmann} T., 2000, \apj, 533, 682

\bibitem[{Cardelli {et~al}\mbox{.}(1989)Cardelli, Clayton, \&
  Mathis}]{cardelli+89}
Cardelli J.~A., Clayton G.~C., Mathis J.~S., 1989, $\backslash$apj, 345, 245

\bibitem[{{Cid Fernandes} {et~al}\mbox{.}(2004){Cid Fernandes}, {Gonz\'{a}lez
  Delgado}, Schmitt, Storchi-Bergmann, Martins, P\'{e}rez, Heckman, Leitherer,
  \& Schaerer}]{cid+04}
{Cid Fernandes} R. {et~al.}, 2004, $\backslash$apj, 605, 105

\bibitem[{{Cid Fernandes} {et~al}\mbox{.}(2005{\natexlab{a}}){Cid Fernandes},
  Mateus, Sodr\'{e}, Stasi$\backslash$'nska, \& Gomes}]{cid+05a}
{Cid Fernandes} R., Mateus A., Sodr\'{e} L., Stasi$\backslash$'nska G., Gomes
  J.~M., 2005{\natexlab{a}}, $\backslash$mnras, 358, 363

\bibitem[{{Cid Fernandes} {et~al}\mbox{.}(2005{\natexlab{b}}){Cid Fernandes},
  Mateus, Sodr\'{e}, Stasi$\backslash$'nska, \& Gomes}]{cid+05b}
{Cid Fernandes} R., Mateus A., Sodr\'{e} L., Stasi$\backslash$'nska G., Gomes
  J.~M., 2005{\natexlab{b}}, $\backslash$mnras, 358, 363

\bibitem[{{Cid Fernandes} {et~al}\mbox{.}(2009){Cid Fernandes}, Schoenell,
  Gomes, Asari, Schlickmann, Mateus, Stasinska, {Sodr\'{e} Jr.},
  Torres-Papaqui, \& {Seagal Collaboration}}]{cid+09}
{Cid Fernandes} R. {et~al.}, 2009, in Revista Mexicana de Astronomia y
  Astrofisica, vol. 27, Vol.~35, Revista Mexicana de Astronomia y Astrofisica
  Conference Series, pp. 127--132

\bibitem[{Coziol {et~al}\mbox{.}(2001)Coziol, Doyon, \& Demers}]{coziol+01}
Coziol R., Doyon R., Demers S., 2001, Monthly Notices of the Royal Astronomical
  Society, 325, 1081

\bibitem[{Coziol {et~al}\mbox{.}(1998)Coziol, Torres, Quast, Contini, \&
  Davoust}]{coziol+98}
Coziol R., Torres C.~A.~O., Quast G.~R., Contini T., Davoust E., 1998,
  $\backslash$apjs, 119, 239

\bibitem[{Cushing {et~al}\mbox{.}(2004)Cushing, Vacca, \& Rayner}]{cushing+04}
Cushing M.~C., Vacca W.~D., Rayner J.~T., 2004, Publications of the
  Astronomical Society of the Pacific, 116, 362

\bibitem[{{Dannerbauer} {et~al}\mbox{.}(2005){Dannerbauer}, {Rigopoulou},
  {Lutz}, {Genzel}, {Sturm}, \& {Moorwood}}]{dannerbauer+05}
{Dannerbauer} H., {Rigopoulou} D., {Lutz} D., {Genzel} R., {Sturm} E.,
  {Moorwood} A.~F.~M., 2005, \aap, 441, 999

\bibitem[{Engelbracht {et~al}\mbox{.}(1998)Engelbracht, Rieke, Rieke, Kelly, \&
  Achtermann}]{engelbracht+98}
Engelbracht C.~W., Rieke M.~J., Rieke G.~H., Kelly D.~M., Achtermann J.~M.,
  1998, $\backslash$apj, 505, 639

\bibitem[{Fischera {et~al}\mbox{.}(2003)Fischera, Dopita, \&
  Sutherland}]{Fishera+03}
Fischera J., Dopita M.~A., Sutherland R.~S., 2003, $\backslash$apjl, 599, L21

\bibitem[{Goldader {et~al}\mbox{.}(1997)Goldader, Goldader, Joseph, Doyon, \&
  Sanders}]{goldader+97}
Goldader J.~D., Goldader D.~L., Joseph R.~D., Doyon R., Sanders D.~B., 1997,
  $\backslash$aj, 113, 1569

\bibitem[{Gu {et~al}\mbox{.}(2006)Gu, Melnick, {Cid Fernandes}, Kunth,
  Terlevich, \& Terlevich}]{gu+06}
Gu Q., Melnick J., {Cid Fernandes} R., Kunth D., Terlevich E., Terlevich R.,
  2006, $\backslash$mnras, 366, 480

\bibitem[{Ho {et~al}\mbox{.}(1995)Ho, Filippenko, \& Sargent}]{ho+95}
Ho L.~C., Filippenko A.~V., Sargent W.~L., 1995, The Astrophysical Journal
  Supplement Series, 98, 477

\bibitem[{Ho {et~al}\mbox{.}(1997)Ho, Filippenko, \& Sargent}]{ho+97}
Ho L.~C., Filippenko A.~V., Sargent W. L.~W., 1997, The Astrophysical Journal
  Supplement Series, 112, 315

\bibitem[{Ilbert {et~al}\mbox{.}(2010)Ilbert, Salvato, {Le Floc'h}, Aussel,
  Capak, McCracken, Mobasher, Kartaltepe, Scoville, Sanders, Arnouts, Bundy,
  Cassata, Kneib, Koekemoer, {Le F\`{e}vre}, Lilly, Surace, Taniguchi, Tasca,
  Thompson, Tresse, Zamojski, Zamorani, \& Zucca}]{ilbert+10}
Ilbert O. {et~al.}, 2010, $\backslash$apj, 709, 644

\bibitem[{{Ivanov} {et~al}\mbox{.}(2000){Ivanov}, {Rieke}, {Groppi},
  {Alonso-Herrero}, {Rieke}, \& {Engelbracht}}]{ivanov+2000}
{Ivanov} V.~D., {Rieke} G.~H., {Groppi} C.~E., {Alonso-Herrero} A., {Rieke}
  M.~J., {Engelbracht} C.~W., 2000, \apj, 545, 190

\bibitem[{{Kennicutt Jr.}(1988)}]{kennicutt88}
{Kennicutt Jr.} R.~C., 1988, $\backslash$apj, 334, 144

\bibitem[{{Kennicutt Jr.}(1992)}]{kennicutt92}
{Kennicutt Jr.} R.~C., 1992, $\backslash$apj, 388, 310

\bibitem[{Kotilainen {et~al}\mbox{.}(2012)Kotilainen, Hyv\"{o}nen, Reunanen, \&
  Ivanov}]{kotilainen+12}
Kotilainen J.~K., Hyv\"{o}nen T., Reunanen J., Ivanov V.~D., 2012, Monthly
  Notices of the Royal Astronomical Society, 425, 1057

\bibitem[{{Larkin} {et~al}\mbox{.}(1998){Larkin}, {Armus}, {Knop}, {Soifer}, \&
  {Matthews}}]{larkin+98}
{Larkin} J.~E., {Armus} L., {Knop} R.~A., {Soifer} B.~T., {Matthews} K., 1998,
  \apjs, 114, 59

\bibitem[{{Lonsdale} {et~al}\mbox{.}(1984){Lonsdale}, {Persson}, \&
  {Matthews}}]{lonsdale+84}
{Lonsdale} C.~J., {Persson} S.~E., {Matthews} K., 1984, \apj, 287, 95

\bibitem[{Maraston(2005)}]{maraston05}
Maraston C., 2005, Monthly Notices of the Royal Astronomical Society, 362, 799

\bibitem[{{Maraston} \& {Str{\"o}mb{\"a}ck}(2011)}]{maraston+11}
{Maraston} C., {Str{\"o}mb{\"a}ck} G., 2011, \mnras, 418, 2785

\bibitem[{{Marigo} {et~al}\mbox{.}(2008){Marigo}, {Girardi}, {Bressan},
  {Groenewegen}, {Silva}, \& {Granato}}]{marigo+08}
{Marigo} P., {Girardi} L., {Bressan} A., {Groenewegen} M.~A.~T., {Silva} L.,
  {Granato} G.~L., 2008, \aap, 482, 883

\bibitem[{Martins {et~al}\mbox{.}(2010)Martins, Riffel, Rodr\'{\i}guez-Ardila,
  Gruenwald, \& de~Souza}]{martins+10}
Martins L.~P., Riffel R., Rodr\'{\i}guez-Ardila A., Gruenwald R., de~Souza R.,
  2010, Monthly Notices of the Royal Astronomical Society, 406, 2185

\bibitem[{Martins {et~al}\mbox{.}(2013)Martins, Rodr\'{\i}guez-Ardila, Diniz,
  Gruenwald, \& de~Souza}]{martins+13}
Martins L.~P., Rodr\'{\i}guez-Ardila A., Diniz S., Gruenwald R., de~Souza R.,
  2013, Monthly Notices of the Royal Astronomical Society

\bibitem[{Mateus {et~al}\mbox{.}(2006)Mateus, Sodr\'{e}, {Cid Fernandes},
  Stasi$\backslash$'nska, Schoenell, \& Gomes}]{mateus+06}
Mateus A., Sodr\'{e} L., {Cid Fernandes} R., Stasi$\backslash$'nska G.,
  Schoenell W., Gomes J.~M., 2006, $\backslash$mnras, 370, 721

\bibitem[{Origlia {et~al}\mbox{.}(1993)Origlia, Moorwood, \&
  Oliva}]{origlia+93}
Origlia L., Moorwood A.~F.~M., Oliva E., 1993, $\backslash$aap, 280, 536

\bibitem[{{Pickles}(1998)}]{pickles98}
{Pickles} A.~J., 1998, \pasp, 110, 863

\bibitem[{Rayner {et~al}\mbox{.}(2003)Rayner, Toomey, Onaka, Denault,
  Stahlberger, Vacca, Cushing, \& Wang}]{rayner+03}
Rayner J.~T., Toomey D.~W., Onaka P.~M., Denault A.~J., Stahlberger W.~E.,
  Vacca W.~D., Cushing M.~C., Wang S., 2003, Publications of the Astronomical
  Society of the Pacific, 115, 362

\bibitem[{Reunanen {et~al}\mbox{.}(2002)Reunanen, Kotilainen, \&
  Prieto}]{reunanen+02}
Reunanen J., Kotilainen J.~K., Prieto M.~A., 2002, Monthly Notices of the Royal
  Astronomical Society, 331, 154

\bibitem[{Reunanen {et~al}\mbox{.}(2003)Reunanen, Kotilainen, \&
  Prieto}]{reunanen+03}
Reunanen J., Kotilainen J.~K., Prieto M.~A., 2003, Monthly Notices of the Royal
  Astronomical Society, 343, 192

\bibitem[{{Reunanen} {et~al}\mbox{.}(2007){Reunanen}, {Tacconi-Garman}, \&
  {Ivanov}}]{reunanen+07}
{Reunanen} J., {Tacconi-Garman} L.~E., {Ivanov} V.~D., 2007, \mnras, 382, 951

\bibitem[{{Riffel} {et~al}\mbox{.}(2011{\natexlab{a}}){Riffel}, {Bonatto}, {Cid
  Fernandes}, {Pastoriza}, \& {Balbinot}}]{riffelRogerio+11}
{Riffel} R., {Bonatto} C., {Cid Fernandes} R., {Pastoriza} M.~G., {Balbinot}
  E., 2011{\natexlab{a}}, \mnras, 411, 1897

\bibitem[{Riffel {et~al}\mbox{.}(2009)Riffel, Pastoriza, Rodr\'iguez-Ardila, \&
  Bonatto}]{riffel+09}
Riffel R., Pastoriza M.~G., Rodr\'iguez-Ardila A., Bonatto C., 2009, Monthly
  Notices of the Royal Astronomical Society, 400, 273

\bibitem[{{Riffel} {et~al}\mbox{.}(2007){Riffel}, {Pastoriza},
  {Rodr{\'{\i}}guez-Ardila}, \& {Maraston}}]{riffel+07}
{Riffel} R., {Pastoriza} M.~G., {Rodr{\'{\i}}guez-Ardila} A., {Maraston} C.,
  2007, \apjl, 659, L103

\bibitem[{Riffel {et~al}\mbox{.}(2008)Riffel, Pastoriza, Rodr\'{\i}guez-Ardila,
  \& Maraston}]{riffel+08}
Riffel R., Pastoriza M.~G., Rodr\'{\i}guez-Ardila A., Maraston C., 2008,
  Monthly Notices of the Royal Astronomical Society, 388, 803

\bibitem[{{Riffel} {et~al}\mbox{.}(2011{\natexlab{b}}){Riffel}, {Riffel},
  {Ferrari}, \& {Storchi-Bergmann}}]{riffel+11}
{Riffel} R., {Riffel} R.~A., {Ferrari} F., {Storchi-Bergmann} T.,
  2011{\natexlab{b}}, \mnras, 416, 493

\bibitem[{{Riffel} {et~al}\mbox{.}(2010){Riffel}, {Storchi-Bergmann}, {Riffel},
  \& {Pastoriza}}]{riffel+10}
{Riffel} R.~A., {Storchi-Bergmann} T., {Riffel} R., {Pastoriza} M.~G., 2010,
  \apj, 713, 469

\bibitem[{{Sanders} \& {Mirabel}(1996)}]{sanders+96}
{Sanders} D.~B., {Mirabel} I.~F., 1996, \araa, 34, 749

\bibitem[{Schlafly \& Finkbeiner(2011)}]{schlafly+11}
Schlafly E.~F., Finkbeiner D.~P., 2011, $\backslash$apj, 737, 103

\bibitem[{{Storchi-Bergmann} {et~al}\mbox{.}(2012){Storchi-Bergmann}, {Riffel},
  {Riffel}, {Diniz}, {Borges Vale}, \& {McGregor}}]{storchi-bergmann+12}
{Storchi-Bergmann} T., {Riffel} R.~A., {Riffel} R., {Diniz} M.~R., {Borges
  Vale} T., {McGregor} P.~J., 2012, \apj, 755, 87

\bibitem[{{Surace} {et~al}\mbox{.}(2004){Surace}, {Sanders}, \&
  {Mazzarella}}]{surace+04}
{Surace} J.~A., {Sanders} D.~B., {Mazzarella} J.~M., 2004, \aj, 127, 3235

\bibitem[{{Telesco}(1988)}]{telesco88}
{Telesco} C.~M., 1988, \araa, 26, 343

\bibitem[{{Vacca} {et~al}\mbox{.}(2003){Vacca}, {Cushing}, \&
  {Rayner}}]{vacca+03}
{Vacca} W.~D., {Cushing} M.~C., {Rayner} J.~T., 2003, \pasp, 115, 389

\bibitem[{{Vanzi} \& {Rieke}(1997)}]{vanzi+97}
{Vanzi} L., {Rieke} G.~H., 1997, \apj, 479, 694

\bibitem[{Worthey \& Ottaviani(1997)}]{worthey+97}
Worthey G., Ottaviani D.~L., 1997, $\backslash$apjs, 111, 377

\bibitem[{Zibetti {et~al}\mbox{.}(2013)Zibetti, Gallazzi, Charlot, Pierini, \&
  Pasquali}]{zibetti+12}
Zibetti S., Gallazzi A., Charlot S., Pierini D., Pasquali A., 2013, Monthly
  Notices of the Royal Astronomical Society, 428, 1479

\end{thebibliography}
\end{document}